\documentclass[prodmode,acmtaco]{acmsmall} 

\pdfoutput=1

\usepackage[utf8]{inputenc}
\usepackage[T1]{fontenc}
\usepackage[ruled]{algorithm2e}

\SetAlFnt{\small}
\SetAlCapFnt{\small}
\SetAlCapNameFnt{\small}
\SetAlCapHSkip{0pt}
\IncMargin{-\parindent}

\usepackage{float}
\usepackage[noend]{algpseudocode}
\algrenewcommand\algorithmicfunction{\textbf{when}}
\algrenewcommand\textproc{} 
\algrenewcommand\algorithmicindent{0.75em}
\usepackage{caption}
\usepackage{subcaption}
\usepackage[acronym]{glossaries}
\newacronym{2pl}{2PL}{Two-phase~Locking}
\newacronym{tm}{TM}{Transactional~Memory}
\newacronym{stm}{STM}{Software~Transactional~Memory}
\newacronym{htm}{HTM}{Hardware~Transactional~Memory}
\newacronym{occ}{OCC}{Optimistic~Concurrency~Control}
\newacronym{pcc}{PCC}{Pot~Concurrency~Control}
\newacronym{rot}{ROT}{Rollback-only~Transaction}
\usepackage{tikz}
\usepackage{pgfplots}
\usepgfplotslibrary{groupplots}
\usetikzlibrary{patterns}
\usepackage{booktabs}
\usepackage{todonotes}

\definecolor{darkcandyapplered}{rgb}{0.64, 0.0, 0.0}
\definecolor{darkcoral}{rgb}{0.8, 0.36, 0.27}
\definecolor{darkcyan}{rgb}{0.0, 0.55, 0.55}
\definecolor{darkgoldenrod}{rgb}{0.72, 0.53, 0.04}
\graphicspath{{diagrams/}}

\acmVolume{13}
\acmNumber{4}
\acmArticle{52}
\acmYear{2016}
\acmMonth{12}

\doi{10.1145/3017993}

\issn{0000-00000}

\pdfoutput=1

\begin{document}

\markboth{T. M. Vale et al.}{Pot: Deterministic transactional execution}

\title{Pot: Deterministic transactional execution}
\author{TIAGO M. VALE and JO\~{A}O A. SILVA
    \affil{NOVA LINCS, DI, FCT, Universidade NOVA de Lisboa}
    RICARDO J. DIAS
    \affil{SUSE Linux GmbH and NOVA LINCS}
    JO\~{A}O M. LOUREN\c{C}O
    \affil{NOVA LINCS, DI, FCT, Universidade NOVA de Lisboa}}

\begin{abstract}
This paper presents Pot, a system that leverages the concept of preordered transactions to achieve deterministic multithreaded execution of programs that use Transactional Memory.
Preordered transactions eliminate the root cause of nondeterminism in transactional execution: they provide the illusion of executing in a deterministic serial order, unlike traditional transactions which appear to execute in a nondeterministic order that can change from execution to execution.
Pot uses a new concurrency control protocol that exploits the serialization order to distinguish between fast and speculative transaction execution modes in order to mitigate the overhead of imposing a deterministic order.
We build two Pot prototypes: one using STM and another using off-the-shelf HTM.
To the best of our knowledge, Pot enables deterministic execution of programs using off-the-shelf HTM for the first time.
An experimental evaluation shows that Pot achieves deterministic execution of~\acrshort{tm} programs with low overhead, sometimes even outperforming nondeterministic executions, and clearly outperforming the state of the art.
\end{abstract}

%
%

%
%



\acmformat{Tiago M. Vale, Jo\~{a}o A. Silva, Ricardo J. Dias, and Jo\~{a}o M. Louren\c{c}o. Pot: Deterministic transactional execution.}

\begin{bottomstuff}
\textbf{New Paper, Not an Extension of a Conference Paper.}

This work is supported by Funda\c{c}\~{a}o para a Ci\^{e}ncia e Tecnologia, Minist\'{e}rio da Ci\^{e}ncia, Tecnologia, e Ensino Superior, under
grants SFRH/BD/84497/2012 and PEst/UID/CEC/04516/2013.

Authors' addresses: T. Vale, J. Silva, R. Dias, {and} J. Louren\c{c}o, Departamento de Inform\'{a}tica, FCT/UNL, Quinta da Torre, 2829-516 Caparica, Portugal.
\end{bottomstuff}

\maketitle

\section{Introduction} 
\label{sec:introduction}
Over the last decade, \acrfull{tm}~\cite{htm-isca-1993,stm-1997} emerged as a viable mechanism to synchronize concurrent accesses to shared state due to an interesting trade-off between ease of use and performance.
With \acrshort{tm}, programmers specify which portions of code should be atomic~(transactions) \emph{without} worrying how to enforce such atomicity.
A concurrency control protocol~(implemented either in software~(\acrshort{stm}), hardware~(\acrshort{htm}), or a mixture of both) enforces atomicity at runtime, providing the illusion that transactions execute one at a time.
\acrshort{tm}~is becoming mainstream, as processors from Intel and IBM already provide support for~\acrshort{htm}~\cite{tm-powerpc-isca-2013,tsx-sc-2013}, the GCC has experimental support for~\acrshort{tm} (using either~\acrshort{stm} or~\acrshort{htm})~\cite{gcc-tm}, and there is ongoing work in integrating~\acrshort{tm} language constructs in C/C++~\cite{cpp-tm-iso}.

Although~\acrshort{tm}~provides a simple programming model it inherits the nondeterministic behavior of multithreaded execution.
Specifically, the order in which transactions appear to execute depends on the nondeterministic interleavings of threads at runtime, so different executions of the same program with the same inputs can yield different outcomes.
In this work we focus on building a TM system that ensures that data race-free programs execute according to a deterministic transaction serialization order.\footnote{This property is known as weak determinism~\cite{kendo-asplos-2009}.}

\begin{figure}[tb]
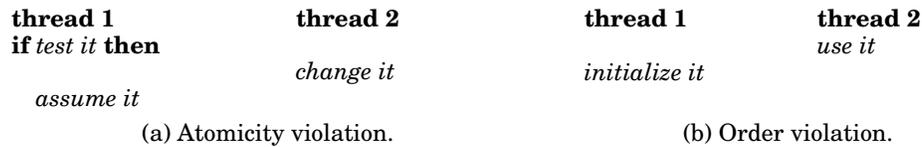

    \raggedright
    \begin{subfigure}[b]{.54\linewidth}
        \begin{minipage}{.49\linewidth}
            \begin{algorithmic}[0]
                \small
                \State \textbf{thread 1}
                \State \textbf{if} \textit{test it} \textbf{then}
                    \Statex
                    \State\hspace{\algorithmicindent} \textit{assume it}
            \end{algorithmic}
        \end{minipage}
        \begin{minipage}{.49\linewidth}
            \begin{algorithmic}[0]
                \small
                \State \textbf{thread 2}
                \Statex
                \State \textit{change it}
                \Statex
            \end{algorithmic}
        \end{minipage}
        \caption{Atomicity violation.}
        \label{subfig:problem_example_atomicity}
    \end{subfigure}
    \raggedleft
    \begin{subfigure}[b]{.44\linewidth}
        \begin{minipage}{.49\linewidth}
            \begin{algorithmic}[0]
                \small
                \State \textbf{thread 1}
                \Statex
                \State \textit{initialize it}
                \Statex
            \end{algorithmic}
        \end{minipage}
        \begin{minipage}{.49\linewidth}
            \begin{algorithmic}[0]
                \small
                \State \textbf{thread 2}
                \State \textit{use it}
                \Statex
                \Statex
            \end{algorithmic}
        \end{minipage}
        \caption{Order violation.}
        \label{subfig:problem_example_order}
    \end{subfigure}
    \caption{
        Example of the most common concurrency bugs~[Lu et al. 2008], with transactions in \textit{italic}.
        In~(a)~the assumption of a predicate is not atomic with its test.
        In~(b)~thread~2 uses a resource before thread~1 initializes it.
    }
    \label{fig:problem_example}
\end{figure}

Having a system that ensures a deterministic transaction serialization order has at least two benefits: (1)~we can execute multiple replicas of a multithreaded application for fault tolerance~\cite{smr-surveys-1990}, and (2)~it helps debugging, or prevents, the most common concurrency bugs~\cite{concurrencybugs-asplos-2008}.
Executing multiple replicas for fault tolerance relies on the assumption that correct replicas always yield the same outputs.
With a deterministic transaction serialization order this assumption is \emph{not} broken under multithreaded execution, so replicas do \emph{not} need to fall back to sequential execution to ensure correctness.
Consequently, replicas potentially make better use of the available resources such as multicore processors.
Regarding concurrency bugs, Fig.~\ref{fig:problem_example} depicts the two most common concurrency bugs~(amounting to 97\% of the non-deadlock bugs) found in a study of 4 real-world applications~\cite{concurrencybugs-asplos-2008}, with transactions highlighted in \textit{italic}.
Fig.~\ref{subfig:problem_example_atomicity} shows an example of an atomicity violation.
Thread~1 tests some predicate, and then executes code that assumes that it is true.
Thread~2 executes code that changes the~predicate's outcome.
If thread~2 interleaves thread~1 after the predicate test, but before the ``then branch,'' thread~1 will execute code that assumes the predicate is true while it is not, which can result in unexpected behavior.
%
%
Fig.~\ref{subfig:problem_example_order} shows an example of an order violation.
Thread~1 initializes some resource that thread~2 uses, but at runtime thread~2 attempts to use the resource before thread~1 initializes it.
These concurrency errors are sensitive to thread interleavings, and in the particular case of \acrshort{tm}, only manifest themselves in particular transaction serialization orders.
Since the transaction serialization order is nondeterministic, the errors are difficult to reproduce and debug.
With a deterministic transaction serialization order, the aforementioned errors either manifest themselves in every execution, or not at all, greatly simplifying the developer's work.

%

In this paper we present Pot, a system that enables deterministic multithreading of \acrshort{tm}-based applications.
While existing work~\cite{destm-pact-2014} also ensures a deterministic transaction serialization order of TM-based applications, Pot: (1)~performs better, which is important when executing multiple replicas for fault tolerance, (2)~is equally helpful when dealing with the most common concurrency bugs such as atomicity and order violations, and (3)~is applicable to both STM and HTM.

In principle, lock-based deterministic multithreading techniques~\cite{coredet-asplos-2010,dthreads-sosp-2011,kendo-asplos-2009,rfdet-ppopp-2014,parrot-sosp-2013} could be used to achieve deterministic execution of~\acrshort{stm} programs~(if, and only if, the~\acrshort{stm} concurrency control protocol is implemented using deterministic locks).
However, such an approach has several drawbacks:~(a)~it cannot be applied to \acrshort{htm}, because the concurrency control is implemented in the hardware,~(b)~it fails to exploit the semantics of transactions to reduce the overhead of ensuring determinism, because determinism is enforced with locks, which are at a level of abstraction lower than transactions, and~(c)~many practical \acrshort{stm}s directly use atomic primitives such as compare-and-swap rather than locks.
Instead, Pot uses the concept of \emph{preordered~transactions} as a principled approach to ensure a deterministic transaction serialization order.
While traditional transactions provide the illusion of executing one at a time in \emph{any} order, preordered~transactions appear to execute in a specific, predefined, order.

To realize preordered~transactions, Pot must address two key challenges:~(1)~guarantee that the predefined serial order is the same across executions, and~(2)~that the outcome of executing transactions is as if they executed serially in the predefined order.
To ensure~(1), Pot's~\emph{sequencer} assigns a sequence number to each new transaction.
The sequence number reflects the transaction's place in a deterministic transaction serialization order.
To ensure~(2)~efficiently, Pot executes transactions concurrently and relies on a new concurrency control protocol that guarantees that the outcome is equivalent to the order defined by the sequencer.
Pot's concurrency control protocol relies on two key techniques:~\emph{ordered~commits} and~\emph{transaction~modes}.
Ordered commits force transactions to commit according to the predefined serialization order.
Transaction modes leverage the key insight that, at any given time, there is always one transaction that is ``the next allowed to commit.''
Pot's concurrency control protocol executes that particular transaction as fast as possible, with virtually no concurrency control overhead~(\emph{fast~mode}) while executing the other transactions using regular mechanisms to maintain correctness in the presence of the fast-mode transaction~(\emph{speculative~mode}).

We built two Pot prototypes, one using \acrshort{stm} and another using off-the-shelf \acrshort{htm}, and evaluate them with the popular STAMP benchmark suite~\cite{stamp-iiswc-2008} and STMBench7~\cite{stmbench7-eurosys-2007}.
Our Pot~\acrshort{stm} implementation clearly outperforms the state of the art in~\acrshort{stm}-based deterministic execution while simultaneously achieving deterministic execution with low overhead, providing promising evidence that using both~\acrshort{stm} and determinism to ease multithreaded programming may be practical.
To the best of our knowledge, Pot also advances the state of the art by enabling deterministic execution of off-the-shelf~\acrshort{htm}-based multithreaded programs for the first time.

The rest of the paper is structured as follows.
\S\ref{sec:design} presents Pot's design, namely its sequencer~(\S\ref{sub:design_ordering}) and concurrency control protocol~(\S\ref{sub:design_execution});
\S\ref{sec:implementation} highlights the challenges and details our implementation of Pot in an STM and an off-the-shelf HTM system; \S\ref{sec:evaluation} reports an experimental evaluation of Pot; we discuss the related work and conclude the paper in \S\ref{sec:related}~and~\S\ref{sec:conclusion}.
\section{Design} 
\label{sec:design}
The standard \acrfull{tm} correctness criterion is opacity~\cite{opacity-ppopp-2008}.
%
%
Traditional concurrency control protocols used to implement opaque transactions, such as \acrlong{2pl}~\cite{concurrencycontrol-1987} or \acrlong{occ}~\cite{occ-1981}, embrace opacity's flexibility and perform two tasks simultaneously while transactions are executing:~(a)~they compute the transaction serialization order~(ordering), and~(b)~control the concurrent execution of transactions to respect that serialization order~(concurrency control).
Since ordering is intertwined with concurrency control, the final transaction serialization order depends on the nondeterministic interleavings that occur at runtime between transactions and thus varies from one execution to the next.
We refer to this execution model as \emph{traditional transactions}.

With \emph{preordered transactions} the serialization order is independent of the interleavings that may occur between transactions because, unlike traditional transactions, preordered transactions already have a place in the serialization order before they are executed.
Conceptually, preordered transactions have a two-phase execution model:~(1)~the \emph{ordering phase} which defines every transactions' place in the serialization order,~and~(2)~the \emph{execution phase} where transactions execute concurrently in such a way that the outcome is equivalent to their sequential execution in the predefined order.
Traditional concurrency control protocols \emph{cannot} be used in the execution phase, because they implement both ordering and concurrency control.
This paper proposes a novel concurrency control protocol that can be used in the execution phase~(\S\ref{sub:design_execution}).
\subsection{Ordering phase: Pot sequencer} 
\label{sub:design_ordering}
A consequence of decoupling ordering and concurrency control is that both the ordering and execution phase, where concurrency control occurs, can be performed separately by two different components.
Ordering is performed by a \emph{sequencer} component that computes some total order over the set of all transactions.

At first glance it seems that the sequencer needs to know which transactions will execute ahead of time, but we can devise generic sequencers that compute the transaction order on-the-fly by defining an order over the application threads and deriving the transaction order from it.
For example, take threads~$t$ and~$u$, with transactions~${(a;b;c)}$ and~${(d;e;f)}$ in their code, respectively.
Consider a sequencer that orders threads using a round-robin scheme, i.e.~${(t;u)}$.
This sequencer defines the transaction order~${(a;d;b;e;c;f)}$.
Now consider that thread~$t$ only executes transaction~$c$ depending on some condition.
The condition may be defined over global state, thread-private state, or a mixture of both.
If the condition is over global state, the respective state must have been read within a transaction, e.g. transaction~$b$, so the condition is always tested over the state resulting from the order~${(a;d;b)}$, yielding a deterministic result.\footnote{\small Assuming the only source of nondeterminism is the transaction serialization order. Techniques to deal with other sources, e.g. randomness, are complementary to this work.}
If thread~$t$ decides not to execute transaction~$c$ the order is~${(a;d;b;e;f)}$.
If thread~$t$'s logic is ``execute~$c$ or~$g$'' instead, the order is~${(a;d;b;e;g;f)}$.

The only requirement of a generic sequencer that derives the transaction order from the thread order is that the events of starting and stopping threads must be processed deterministically by the sequencer with respect to the transaction order.
To do so, since transactions appear to execute in a deterministic order, Pot treats thread start/stop events as if they are transactions.
Take threads~$t$ and~$u$, with transactions~${(a;b;c)}$ and~${(d;e;f)}$, respectively, where transaction~$b$ is the creation of a new thread~$v$ with transactions~${(g;h)}$.
If we organize threads in a tree where the main thread is the root, the remaining threads are children of the thread that spawned them, and let the tree's post-order traversal specify the thread order, a round-robin sequencer defines the transaction order~${(a;d;b;e;g;c;f;h)}$.

It is also possible to use application-specific sequencers.
For example, we may record the transaction commit order in a nondeterministic execution and then feed it to a sequencer to replay the recorded execution.
We can also have sequencers that explicitly define a transaction order, e.g. ~${(a;b;c;d;e;f)}$, but these need to take care because if a thread decides not to execute a transaction in the order then the program would hang waiting for it to execute. (We can detect this situation and abort the application with an error.)

Our design works best for workloads in which threads perform transactions regularly.
Optimizing for workloads with very heterogeneous thread behaviors is an open problem left for future work.
\subsection{Execution phase: \acrlong{pcc}} 
\label{sub:design_execution}
Transactions may execute once they go through the ordering phase.
At the core of the execution phase is a concurrency control protocol that guarantees equivalence to the serialization order defined in the ordering phase.
The straightforward way to implement such concurrency control protocol is to simply execute transactions sequentially.
However this approach is clearly suboptimal as it does not take advantage of the inherent parallelism present in today's multicore architectures.

This section describes \acrfull{pcc}, a new protocol that executes transactions concurrently while guaranteeing equivalence to the serial order defined by the sequencer.
We design \acrshort{pcc} by modifying \acrfull{occ}, which works as follows.
An \acrshort{occ} transaction consists of one, or more, speculative executions.
A speculative execution is divided into three phases:~(1)~the read~phase,~(2)~the validation~phase, and~(3)~the write~phase.
The read phase records the objects read by the transaction in the transaction's read~set.
Write operations do not modify the shared state; instead the transaction defers its updates and logs them in its write~set.
Therefore locations that are both read and modified occur in both the read and the write~sets.
After the read~phase, the transaction undergoes a validation~phase where it checks whether any concurrently committed transaction's updates overlap with its read~set.
If so the transaction is aborted to respect opacity, and can be retried; otherwise it proceeds to the next phase.
Finally, the transaction enters the write~phase where it atomically updates all objects in its write~set with the values buffered during the read~phase.

We have chosen \acrshort{occ} as the base for \acrshort{pcc} because \acrshort{occ} is suitable for dynamic transactions, i.e. transactions for which it is very difficult~(or even impossible) to identify their read/write sets in advance.
Dynamic transactions are common in general-purpose \acrshort{tm}-based programs due to aliasing and the unstructured nature of the heap.
In fact, most \acrshort{stm} and all existing \acrshort{htm} concurrency control protocols are optimistic.

Next, we present \acrshort{pcc} incrementally.
First, we describe the baseline \acrshort{occ} protocol in~\S\ref{subsub:design_execution_baseline}, and then present our methodology to transform the baseline \acrshort{occ} protocol into \acrshort{pcc} by applying two key techniques: ordered commits, in~\S\ref{subsub:design_execution_commits}, and transaction modes, in~\S\ref{subsub:design_execution_modes}.
\subsubsection{Baseline protocol} 
\label{subsub:design_execution_baseline}
\begin{figure*}[tb]
    \centering
    \begin{subfigure}[b]{.30\textwidth}
        \begin{algorithmic}[1]
            \footnotesize
            \Function{txn\_start}{$t$}
                \State
            \EndFunction
            \Function{txn\_write}{$t, o, v$}
                \State \Call{deferred\_update}{$o, v, W_t$}
            \EndFunction
            \Function{txn\_read}{$t, o$}
                \State \Call{consistent\_read}{$o, R_t, W_t$} \textbf{or} abort
            \EndFunction
            \Function{txn\_commit}{$t$}
                \State \textbf{atomically}
                    \State\hspace{\algorithmicindent} \textbf{if} \Call{validate}{$R_t$}
                    \State\hspace{\algorithmicindent}\hspace{\algorithmicindent} \Call{writeback}{$W_t$}
                    \State\hspace{\algorithmicindent} \textbf{else} 
                        \State\hspace{\algorithmicindent}\hspace{\algorithmicindent} abort
            \EndFunction
        \end{algorithmic}
        \caption{\acrshort{occ}.}
        \label{subfig:methodology_occ}
    \end{subfigure}
    \begin{subfigure}[b]{.34\textwidth}
        \begin{algorithmic}[1]
            \footnotesize
            \Function{txn\_start}{$t, sn$}
                \State $sn_t \gets sn$
            \EndFunction
            \Function{txn\_write}{$t, o, v$}
                \State \Call{deferred\_update}{$o, v, W_t$}
            \EndFunction
            \Function{txn\_read}{$t, o$}
                \State \Call{consistent\_read}{$o, R_t, W_t$} \textbf{or} abort
            \EndFunction
            \Function{txn\_commit}{$t$}
                \State \textbf{wait until} $sn_c =$ \Call{pred}{$sn_t$}
                \State \textbf{if} \Call{validate}{$R_t$}
                    \State\hspace{\algorithmicindent} \Call{writeback}{$W_t$}
                    \State\hspace{\algorithmicindent} $sn_c \gets sn_t$
                \State \textbf{else}
                    \State\hspace{\algorithmicindent} abort
            \EndFunction
        \end{algorithmic}
        \caption{Speculative \acrshort{pcc}.}
        \label{subfig:methodology_pcc_spec}
    \end{subfigure}
    \begin{subfigure}[b]{.32\textwidth}
        \begin{algorithmic}[1]
            \footnotesize
            \Function{$sn_c = $ pred}{$sn_t$}
                \State \textbf{if} \Call{validate}{$R_t$}
                    \State\hspace{\algorithmicindent} \Call{writeback}{$W_t$}
                \State \textbf{else}
                    \State\hspace{\algorithmicindent} abort
            \EndFunction
            \Function{txn\_write}{$t, o, v$}
                \State \Call{direct\_update}{$o, v$}
            \EndFunction
            \Function{txn\_read}{$t, o$}
                \State \Call{read}{$o$}
            \EndFunction
            \Function{txn\_commit}{$t$}
                \State $sn_c \gets sn_t$
            \EndFunction
        \end{algorithmic}
        \caption{Fast \acrshort{pcc}.}
        \label{subfig:methodology_pcc_fast}
    \end{subfigure}
    \caption{
        Methodology to transform \acrfull{occ} into \acrfull{pcc}.
        Fig.~\ref{subfig:methodology_occ} models a typical \acrshort{occ} transaction.
        Figs.~\ref{subfig:methodology_pcc_spec}~and~\ref{subfig:methodology_pcc_fast} model a \acrshort{pcc} transaction in speculative and fast mode, respectively.
        $sn_c$~represents the sequence number of the last committed transaction.
        $sn_t$,~$R_t$~and~$W_t$~represent the sequence number, read~set, and write~set of transaction~$t$, respectively.
    }
    \label{fig:methodology}
\end{figure*}
Consider the protocol depicted in Fig.~\ref{subfig:methodology_occ}, modeling a typical \acrshort{occ} scheme~\cite{tl2-disc-2006,occ-1981}.
The read~phase occurs after~$txn\_start$~and before either~$txn\_commit$~or~$txn\_abort$,~and consists of invocations to~$txn\_read$~and/or~$txn\_write$.
Both the validation and write~phase occur during~$txn\_commit$.

\noindent{\emph{Read phase.}} 
Write operations intending to update object~$o$'s value to~$v$, buffer the update in~$W_t$~(Fig.~\ref{subfig:methodology_occ},~line~4).
Read operations on an object~$o$~log the access in the transaction's read~set~$R_t$ and return~(a)~the buffered value for~$o$~in the write~set~$W_t$, if existing, or~(b)~read a value of~$o$~from the shared state consistent with the rest of the read~set~(line~6).
If it is not possible to read a consistent value the transaction aborts.
For example, take two objects~$x$~and~$y$, both initially 0.
Transaction~$t$~observes~$x = 0$.
Meanwhile, another transaction commits and sets both~$x$~and~$y$~to 1.
If transaction~$t$~attempts to read~$y$~it can either return 0 or abort, but it must never return 1, because~$x = 0$~and~$y = 1$~is not possible under opacity.

\noindent{\emph{Validation phase.}} 
The validation~phase iterates the read~set and checks that the observed values are still coherent, i.e., all the observed values remain the same~(line~9).

\noindent{\emph{Write phase.}} 
If validation is successful then transaction~$t$~enters its write~phase and directly~updates the objects in its write~set with the values buffered during the read~phase, creating a new version of the shared state~(line~10).


\noindent{\emph{Correctness.}} 
This protocol guarantees opacity mainly due to the atomicity of the validation and write~phases~(lines~8--12).
If the validation~phase is successful then none of the read objects have been modified since the transaction's read~phase.
This means that the read~phase happens in the same logical instant of the validation~phase.
Since the validation and write~phase occur atomically, the write~phase also happens in the same logical instant of the read~phase.
Therefore, transaction~$t$~appears to have been the sole transaction executing.
Hence~$t$~is serialized after all the transactions that wrote the values~$t$~observed, and before any transactions that eventually observe the values~$t$~wrote.
\subsubsection{Ordered commits} 
\label{subsub:design_execution_commits}
The \acrshort{occ} protocol described in the previous section provides the illusion that transactions execute one at a time.
However, the order in which transactions appear to execute is not deterministic because it depends on the interleavings between transactions' operations that will occur at runtime. 

To adhere to the serial order predefined in the ordering~phase, we make two key observations:~(a)~\acrshort{occ} transactions only modify shared state during their write~phase, and~(b)~each transactions' place in the serialization order depends on the relative order in which each transaction~(atomically) performs its validation and write~phase.
If we restrict transactions to execute their validation and write~phases in the order defined by the sequencer, we guarantee that the outcome is equivalent to the respective ordered sequential execution.

To transform the \acrshort{occ} protocol described in the previous section into \acrshort{pcc}, we start by updating the~$txn\_start$~operation to have an additional parameter, a sequence number~$sn$,~that reflects the order of transaction~$t$~in the serialization order defined by the sequencer~(Fig.~\ref{subfig:methodology_pcc_spec},~line~1).
Transaction~$t$~is preordered after the transaction with sequence number ${predecessor(sn_t)}$ and before the transaction with sequence number ${successor(sn_t)}$.
We force transactions to commit according to the predefined order by inserting a conditional wait in $txn\_commit$.
When transaction~$t$~wants to commit, it waits until the transaction with sequence number~$predecessor(sn_t)$~commits~(line~8).
To this end, transactions communicate via a~$sn_c$~object whose value is the sequence number of the last committed transaction~(line~11).

\noindent{\emph{Correctness.}} 
In the original \acrshort{occ} protocol correctness is guaranteed by atomically executing both the validation and write~phase.
However, the order in which active transactions execute those phases depends on their nondeterministic multithreaded execution.
To conform with the predefined order the atomic block is replaced with a conditional wait that restricts the order in which transactions are allowed to commit.
Specifically, a transaction~$t$~that finishes its read~phase is only allowed to perform the validation and write~phases \emph{after} the transaction that directly precedes~$t$~in the serial order has completed.
Since transactions are totally ordered, \emph{only one transaction at a time} can escape the conditional waiting on line~8.
Correctness is maintained because the conditional wait also guarantees atomicity.
The atomicity scope is between the wait condition~(line~8) and updating~$sn_c$~(line~11).
\subsubsection{Transaction modes} 
\label{subsub:design_execution_modes}
\acrshort{occ} employs a set of techniques to guarantee correctness, such as read and write~sets, read~set validation and deferred~updates.
With \acrshort{occ} \emph{all} transactions are executed using the aforementioned techniques because \emph{any} transaction \emph{may} become the next transaction in the serialization order, which is being defined as transactions execute.
Using such techniques imposes additional overhead when compared with an execution without any concurrency control.

However, unlike in \acrshort{occ}, in \acrshort{pcc} the serialization order is predefined.
Since \acrshort{pcc} restricts the order in which transactions commit, they may now have to wait for their turn to commit, leading to a loss of parallelism.
To mitigate this loss of parallelism, we make the key observation that at any moment there is always a single transaction, which we refer to as \emph{fast}, which is the next transaction that is allowed to commit.
We exploit the fact that the fast transaction is the next transaction allowed to commit to execute it without most concurrency control overheads. 
Hence, we distinguish between two types of transactions: fast and speculative.
We describe both fast and speculative modes below.

\noindent{\textbf{Fast transaction.}} 
A fast transaction~$t$~is the \emph{only} active transaction whose predecessors are all completed.
A fast transaction is the next, and only, transaction allowed to commit.
It can be executed more efficiently by merging the read and write~phases and completely removing the validation~phase, thus eschewing most of the traditional \acrshort{occ} techniques and associated overhead.
Fast transactions execute according to the protocol in~Fig.~\ref{subfig:methodology_pcc_fast}.

\noindent{\emph{Read phase.}} 
Write operations no longer perform deferred~updates; instead they use direct~updates~(line~7).
Since updates are installed in place during the now combined read-write phase, read operations are reduced to simply reading the current object's value with no additional consistency checks or read~set tracking~(line~9).

\noindent{\emph{Validation phase.}} 
Fast transactions are guaranteed to execute to completion without interference from other active transactions, thus the validation~phase is unnecessary.
(Transactions that switch on~the~fly to fast~mode need to validate the speculative execution done up to that point; we elaborate below.)

\noindent{\emph{Write phase.}} 
The write phase is implicitly executed during the read~phase due to the direct~update strategy, therefore the ``write~back'' step is also completely eliminated.
%

\noindent{\emph{Correctness.}} 
Our argument for correctness is the same as for the ordered~commits technique.
However a fast transaction does not speculatively perform the read~phase and wait for its turn to transition to the validation and write~phases.
Instead the fast transaction executes the now combined read-write~phase when it is already its turn to commit.
A fast transaction is effectively given exclusive write permission to the shared state until it commits, so merging the read and write phases by replacing deferred with direct~updates, and removing the validation~phase, does not affect correctness.

\noindent{\textbf{Speculative transaction.}} 
A transaction whose turn to commit has not yet come is a speculative transaction, and it follows the ordered~commit protocol~(\S\ref{subsub:design_execution_commits}).

\noindent{\textbf{Live promotion.}} 
Since fast transactions bypass most concurrency control overhead, a live speculative transaction~$t$, i.e. still executing its read~phase, immediately switches to fast~mode as soon as ${sn_c = predecessor(sn_t)}$ holds~(line~1).
Upon a live promotion, transaction~$t$~eagerly validates the portion of the read~phase it has executed so far~(line~2).
If the validation is successful then~$t$~applies any pending writes to the shared state, \emph{without} updating~$sn_c$, and executes its remaining operations in fast mode~(line~3).
Otherwise~$t$~aborts and retries in fast mode~(line~5).

\noindent{\textbf{Explicit aborts.}} 
If the transaction API has an explicit $txn\_abort$ operation to abort the current transaction, fast transactions must keep the write~set as an undo~log, i.e. remember the values they overwrite to restore them upon abort.
The $txn\_abort$ operation may allow the developer to specify a ``no~retry'' policy, i.e. abort the transaction without retrying it afterwards.
If so, these ``no~retry'' aborts must comply with the predefined order as they are equivalent to committing the current transaction as read~only.
This is done by processing a ``no~retry'' explicit abort as a commit.
For example, a speculative~tansaction waits for its turn, validates its read~set, and updates $sn_c$ if validation is sucessful, or retries if not.
A fast~transaction restores the write~set~(undo~log) and updates $sn_c$.

\noindent{\textbf{Multiple simultaneous fast transactions.}} 
Multiple fast transactions can safely execute in parallel given additional knowledge about transactions.
A string of successive transactions that do not have read-write nor write-write conflicts between themselves can all execute simultaneously as fast transactions, because the final outcome is independent of the order in which they commit.
To implement multiple simultaneous fast transactions the runtime requires a compatibility matrix of all transactions.
When a transaction becomes fast it publishes its information:~transaction identifier, sequence number, and that it is active.
Using this scheme, a transaction knows it can switch to fast mode if:~(1) its predecessor is already fast~(active or finished), and~(2) it is compatible with all currently active fast transactions.
If both conditions hold, the transaction can switch to fast~mode.
%
\section{Implementation} 
\label{sec:implementation}
We implemented a Pot prototype consisting of an implementation of a sequencer and two concurrency control protocols: one where transactions execute using \acrshort{stm} and another where transactions execute using \acrshort{htm}.
Our sequencer implementation is generic and derives the transaction order from a round-robin thread order~(\S\ref{sub:design_ordering}).
Next, we describe our~\acrshort{stm}~(\S\ref{sub:implementation_stm}) and~\acrshort{htm}~(\S\ref{sub:implementation_htm}) implementations.

\subsection{\Acrlong{stm}} 
\label{sub:implementation_stm}
\begin{figure}[tb]
    \centering
    \begin{subfigure}[b]{.37\textwidth}
        \begin{algorithmic}[1]
            \scriptsize
            \Function{txn\_start}{$t$}
                \State $rv_t \gets gv$
                \State acquire-fence
            \EndFunction
            \Function{txn\_write}{$t, addr, val$}
                \State add ($addr, val$) to $W_t$
            \EndFunction
            \Function{txn\_read}{$t, addr$}
                \State \textbf{if} $addr \in W_t$ 
                    \State\hspace{\algorithmicindent} \Return $W_t(addr)$
                \State $v1 \gets$ \Call{get-vlock}{$addr$}
                \State acquire-fence
                \State $value \gets$ \Call{read}{$addr$}
                \State acquire-fence
                \State $v2 \gets$ \Call{get-vlock}{$addr$}
                \State \textbf{if} \Call{unlocked}{$v1$} $\wedge$ $v1 \leq rv_t$ $\wedge$ $v1 = v2$
                    \State\hspace{\algorithmicindent} add $addr$ to $R_t$
                    \State\hspace{\algorithmicindent} \Return {value}
                \State \textbf{else} abort
            \EndFunction
            \Function{txn\_commit}{$t$}
                \State \textbf{for each} $(addr, -) \in W_t$
                    \State\hspace{\algorithmicindent} \textbf{if} \Call{try-lock}{$addr$} fails
                         \State\hspace{\algorithmicindent}\hspace{\algorithmicindent} abort
                \State $wv_t \gets$ \Call{atomic-add-fetch}{$gv, 2$}
                \State \textbf{for each} $addr \in R_t$
                    \State\hspace{\algorithmicindent} $v \gets$ \Call{get-vlock}{$addr$}
                    \State\hspace{\algorithmicindent} \textbf{if} \Call{locked-by-other}{$v$} $\vee$ $v > rv_t$
                        \State\hspace{\algorithmicindent}\hspace{\algorithmicindent} abort
                \State \textbf{for each} $(addr, val) \in W_t$
                    \State\hspace{\algorithmicindent} \Call{write}{$addr, val$}
                \State release-fence
                \State \textbf{for each} $(addr, \_) \in W_t$
                    \State\hspace{\algorithmicindent} \Call{set-and-unlock}{$addr, wv_t$}
            \EndFunction
        \end{algorithmic}
        \caption{Original TL2.}
        \label{subfig:pcc_stm_tl2}
    \end{subfigure}
    \begin{subfigure}[b]{.32\textwidth}
        \begin{algorithmic}[1]
            \scriptsize
            \Function{txn\_start}{$t$}
                \State $rv_t \gets gv$
                \State acquire-fence
                \State \textbf{if} first attempt
                    \State\hspace{\algorithmicindent} $wv_t \gets$ \Call{get-seq-no}{$tid$}
            \EndFunction
            \Function{txn\_write}{$t, addr, val$}
                \State add ($addr, val$) to $W_t$
            \EndFunction
            \Function{txn\_read}{$t, addr$}
                \State \textbf{if} $addr \in W_t$
                    \State\hspace{\algorithmicindent} \Return $W_t(addr)$
                \State $v1 \gets$ \Call{get-version}{$addr$}
                \State acquire-fence
                \State $value \gets$ \Call{read}{$addr$}
                \State acquire-fence
                \State $v2 \gets$ \Call{get-version}{$addr$}
                \State \textbf{if} $v1 \leq rv_t$ $\wedge$ $v1 = v2$
                    \State\hspace{\algorithmicindent} add $addr$ to $R_t$
                    \State\hspace{\algorithmicindent} \Return {value}
                \State \textbf{else} abort
            \EndFunction
            \Function{txn\_commit}{$t$}
                \State \textbf{wait until} $gv = wv_t - 1$
                \State acquire-fence
                \State \textbf{for each} $addr \in R_t$
                    \State\hspace{\algorithmicindent} $v \gets$ \Call{get-version}{$addr$}
                    \State\hspace{\algorithmicindent} \textbf{if} $v > rv_t$
                        \State\hspace{\algorithmicindent}\hspace{\algorithmicindent} abort
                \State \textbf{for each} $(addr, val) \in W_t$
                    \State\hspace{\algorithmicindent} \Call{set-version}{$addr, wv_t$}
                    \State\hspace{\algorithmicindent} release-fence
                    \State\hspace{\algorithmicindent} \Call{write}{$addr, val$}
                \State release-fence
                \State $gv \gets wv_t$
            \EndFunction
        \end{algorithmic}
        \caption{Speculative \acrshort{pcc}.}
        \label{subfig:pcc_stm_spec}
    \end{subfigure}
    \begin{subfigure}[b]{.29\textwidth}
        \begin{algorithmic}[1]
            \scriptsize
            \Function{$gv = wv_t - 1$}{}
                \State acquire-fence
                \State \textbf{for each} $addr \in R_t$
                    \State\hspace{\algorithmicindent} $v \gets$ \Call{get-version}{$addr$}
                    \State\hspace{\algorithmicindent} \textbf{if} $v > rv_t$
                        \State\hspace{\algorithmicindent}\hspace{\algorithmicindent} abort
                \State \textbf{for each} $(addr, val) \in W_t$
                    \State\hspace{\algorithmicindent} \Call{set-version}{$addr, wv_t$}
                    \State\hspace{\algorithmicindent} release-fence
                    \State\hspace{\algorithmicindent} \Call{write}{$addr, val$}
            \EndFunction
            \Function{txn\_write}{$t, addr, val$}
                \State \Call{set-version}{$addr, wv_t$}
                \State release-fence
                \State \Call{write}{$addr, val$}
            \EndFunction
            \Function{txn\_read}{$t, addr$}
                \State \Return~\Call{read}{$addr$}
            \EndFunction
            \Function{txn\_commit}{$t$}
                \State release-fence
                \State $gv \gets wv_t$
            \EndFunction
        \end{algorithmic}
        \caption{Fast \acrshort{pcc}.}
        \label{subfig:pcc_stm_fast}
    \end{subfigure}
    \caption{\acrfull{pcc} \acrshort{stm} implementation.}
    \label{fig:pcc_stm}
\end{figure}
The ordered~commits technique ensures that only one transaction executes its commit procedure at a time.
In NOrec~\cite{norec-ppopp-2010} commits are also sequential.
While this similarity makes NOrec a potential baseline for Pot, NOrec eschews per-memory location metadata and uses value-based validation instead.
Consequently, speculative transactions are unable to identify which particular memory location is written when the fast transaction performs a write.
%
As such, implementing fast transactions while still preserving opacity would require that, every time a fast transaction performs a write, all speculative transactions would have to validate their entire read set, regardless of which specific memory location was written by the fast transaction.
Instead, our Pot \acrshort{stm} protocol is based on TL2~\cite{tl2-disc-2006}, a popular \acrshort{stm} that uses per-memory location metadata, so that speculative transactions do not have to perform incremental validation on reads.
%
%
%

\noindent{\textbf{Baseline STM transaction.}} 
In a nutshell, TL2 works as follows.
There is a global version and a table of versioned locks, i.e., a version and a lock bit implemented as a single value---vlocks for short.
Odd versions are locked and even versions are unlocked.
Each memory address is mapped to one vlock.
When a transaction starts, it samples the global version~$gv$~to~$rv_t$~and performs an acquire fence~(Fig.~\ref{subfig:pcc_stm_tl2},~lines~2--3).
The transaction can safely read any value whose version is less than or equal to its~$rv_t$~sampling.
The fence with acquire semantics ensures that this transaction observes all the memory writes performed by the transaction that updated~$gv$'s~value to~$rv_t$.
Write operations are buffered in the write~set~(line~5).
Read operations return the value of a buffered write if there is any~(line~7--8).
Otherwise, they perform a consistent read by:~(1)~reading the address' vlock to~$v1$~(line~9), (2)~performing an acquire fence~(line~10), (3)~reading the memory address~(line~11), (4)~performing another acquire fence~(line~12),~and (5)~reading the vlock again to~$v2$~(line~13).
The first fence ensures that the memory address value is at least as recent as~$v1$.
(If~$v1$~is 42, then the value read has version 42 or newer.)
The second fence ensures that if the value is newer than~$v1$,
then~$v2$~is at least as recent as the value's version.
(If the value read has version 43,~$v2$~is 43 or newer.)
If~$v1$~is not locked, and~$v1 \leq rv_t$,~and $v1 = v2$,~then the read successfully returns a consistent value; otherwise, the transaction aborts~(lines~14--17).

The commit operation locks every address in the write~set by performing a compare-and-swap on their vlocks. If any of the compare-and-swap operations fails, then the transaction releases any acquired locks and aborts~(lines~19--21).
After successfully acquiring the vlocks, the transaction performs an atomic add-and-fetch by~2 on~$gv$~and stores~$gv$'s~new value in~$wv_t$~(line~22). 
Then, the transaction validates its read~set by checking whether all memory addresses read are unlocked and their version is still compatible with~$rv_t$.
If any check fails then the transaction restores any acquired locks and aborts~(lines~23--26).
Note that the atomic add-and-fetch operation ensures that: (1)~any other transaction that starts meanwhile and observes ${gv = wv_t}$ will at least observe all the write~set vlocks as acquired, and (2)~if any transactions committed since this transaction started, i.e. ${wv_t > rv_t + 2}$, and wrote to a memory address read by this transaction, then the read~set validation will observe vlocks as locked or with a version newer than~$rv_t$. 

At this point the transaction successfully commits.
It writes~back any buffered writes, performs a release fence, and unlocks the write~set, setting every vlock to~$wv_t$.
The release fence ensures that if any transaction observes a vlock with version~$wv_t$~then it also observes the value written by the transaction.

\noindent{\textbf{Speculative STM transaction.}} 
To implement \acrshort{pcc}, we leverage the fact that TL2 uses a global version and retrofit sequence numbers directly as versions.
Thus, transactions communicate the commit order via~$gv$.
A consequence of ordered commits is that we no longer require locks, just versions, as they were only needed due to concurrent commits.

When a transaction starts for the first time, it requests its sequence number~$wv_t$~from the sequencer by supplying the thread's identifier~$tid$~(Fig.~\ref{subfig:pcc_stm_spec},~lines~4--5).
%
%
Read operations are similar to TL2 except that we no longer test if the address is locked~(line~16).
When the transaction attempts to commit, if necessary it waits until ${gv = wv_t - 1}$~(line~21).
Once~${gv = wv_t - 1}$,~we perform an acquire fence that ensures that the following read~set validation observes the newest version of the addresses read~(line~23--26).
In the write~back step, we first update the address' version, perform a release fence, and then write the new value~(lines~27--30).
As discussed before, the release fence ensures that if any transaction observes the written value, it also observes the new version number.
Finally, the transaction updates~$gv$,~signaling the next transaction that it is its turn to commit~(line~32).
The update of $gv$ is preceded by a release fence to ensure that all transactions that see the new value of $gv$ will also see the new values for the objects written in the write~back.

%

\noindent{\textbf{Fast STM transaction.}} 
The fast~mode write operation is equivalent to the write~back step of a speculative transaction, i.e. updates the version number, performs a release fence, and writes the new value~(Fig.~\ref{subfig:pcc_stm_fast},~lines~12--14).
The read operation is reduced to a regular load from memory~(line~16), and the commit operation simply updates~$gv$~(line~19).

\noindent{\textbf{Live promotion.}} 
A speculative STM transaction~$t$ changes to fast on~the~fly when it detects that it is its turn, i.e. ${gv = wv_t - 1}$~(Fig.~\ref{subfig:pcc_stm_fast},~lines~1--10).
In our implementation we check whether the condition holds whenever the speculative transaction begins, reads, or writes.
\subsection{\Acrlong{htm}} 
\label{sub:implementation_htm}
\begin{figure*}[tb]
    \centering
    \begin{subfigure}[b]{.28\textwidth}
        \begin{algorithmic}[1]
            \footnotesize
            \Function{txn\_start}{$t$}
                \State \textbf{if} first attempt
                    \State\hspace{\algorithmicindent} $path_t \gets$ HW
                    \State\hspace{\algorithmicindent} $tries_t \gets 10$
                \State \textbf{wait while} \Call{locked}{$gl$}
                \State \Call{tbegin}{}
                \State \textbf{if} \Call{locked}{$gl$}
					\State\hspace{\algorithmicindent} abort
                \State execute app. code
            \EndFunction
            \Function{txn\_abort}{$t$}
                \State \textbf{if} persistent
                    \State\hspace{\algorithmicindent} $tries_t \gets 0$
                \State \textbf{else}
                    \State\hspace{\algorithmicindent} $tries_t \gets tries_t - 1$
                \State \textbf{if} $tries_t = 0$
                    \State\hspace{\algorithmicindent} \Call{lock}{$gl$}
                    \State\hspace{\algorithmicindent} $path_t \gets$ SW
                    \State\hspace{\algorithmicindent} execute app. code
            \EndFunction
            \Function{txn\_commit}{$t$}
                \State \textbf{if} $path_t = $ HW
                    \State\hspace{\algorithmicindent} \Call{tcommit}{}
                \State \textbf{else}
                    \State\hspace{\algorithmicindent} \Call{unlock}{$gl$}
            \EndFunction
        \end{algorithmic}
        \caption{Standard \acrshort{htm}.}
        \label{subfig:pcc_htm_orig}
    \end{subfigure}
    \begin{subfigure}[b]{.32\textwidth}
        \begin{algorithmic}[1]
            \footnotesize
            \Function{txn\_start}{$t$}
                \State \textbf{if} first attempt
                    \State\hspace{\algorithmicindent} $path_t \gets$ HW
                    \State\hspace{\algorithmicindent} $tries_t \gets 10$
                    \State\hspace{\algorithmicindent} $sn_t \gets$ \Call{get-seq-no}{$tid$}
                \State \textbf{wait while} \Call{locked}{$gl$}
                \State \textbf{if} $sn_c = sn_t$
                    \State\hspace{\algorithmicindent} \Call{tbegin}{$ROT$}
                    \State\hspace{\algorithmicindent} switch to fast mode
                \State \textbf{else}
                    \State\hspace{\algorithmicindent} $sn \gets sn_c$
                    \State\hspace{\algorithmicindent} \Call{tbegin}{}
                    \State\hspace{\algorithmicindent} \textbf{if} \Call{locked}{$gl$}
						\State\hspace{\algorithmicindent}\hspace{\algorithmicindent} abort
                \State execute app. code
            \EndFunction
            \Function{txn\_abort}{$t$}
                \State \textbf{if} persistent 
                    \State\hspace{\algorithmicindent} \textbf{wait until} $sn_c = sn_t - 1$
                \State \textbf{else}
                    \State\hspace{\algorithmicindent} \textbf{wait until} $sn_c > sn$
            \EndFunction
            \Function{txn\_commit}{$t$}
                \State \Call{tsuspend}{}
                \State \textbf{wait until} $sn_c = sn_t$
                \State \Call{tresume}{}
                \State \Call{tcommit}{}
                \State $sn_c \gets sn_t$
            \EndFunction
        \end{algorithmic}
        \caption{Speculative \acrshort{pcc}.}
        \label{subfig:pcc_htm_spec}
    \end{subfigure}
    \begin{subfigure}[b]{.28\textwidth}
        \begin{algorithmic}[1]
            \footnotesize
            \Function{txn\_start}{$t$}
                \State \Call{tbegin}{$ROT$}
                \State execute app. code
            \EndFunction
            \Function{txn\_abort}{$t$}
                \State \textbf{if} persistent
                    \State\hspace{\algorithmicindent} $tries_t \gets 0$
                \State \textbf{else}
                    \State\hspace{\algorithmicindent} $tries_t \gets tries_t - 1$
                \State \textbf{if} $tries_t = 0$
                    \State\hspace{\algorithmicindent} \Call{lock}{$gl$}
                    \State\hspace{\algorithmicindent} $path_t \gets$ SW
                    \State\hspace{\algorithmicindent} execute app. code
            \EndFunction
            \Function{txn\_commit}{$t$}
                \State \textbf{if} $path_t =$ HW 
                    \State\hspace{\algorithmicindent} \Call{tcommit}{}
                \State \textbf{else}
                    \State\hspace{\algorithmicindent} \Call{unlock}{$gl$}
                \State $sn_c \gets sn_t$
            \EndFunction
        \end{algorithmic}
        \caption{Fast \acrshort{pcc}.}
        \label{subfig:pcc_htm_fast}
    \end{subfigure}
    \caption{\acrfull{pcc} \acrshort{htm} implementation.}
    \label{fig:pcc_htm}
\end{figure*}
Implementing \acrshort{pcc} in \acrshort{htm} poses unique challenges when compared with an \acrshort{stm} implementation.
Existing \acrshort{htm}s use the cache to maintain the read and write~set, and rely on the cache coherence protocol to detect conflicts.
\acrshort{htm}s are also \emph{best effort}, i.e., hardware transactions are not guaranteed to eventually commit, even in the absence of conflicts, be it because the transaction's footprint exceeds the cache capacity, or due to the execution of an illegal instruction, an interrupt, a page fault, etc.
Therefore, we must always provide a software fallback to guarantee progress.
These characteristics pose three challenges, namely: (a)~how to ensure that transactions eventually progress, (b)~how to implement ordered commits without inducing false conflicts,~and (c)~how to implement fast transactions.

In our prototype we ensure progress using the most common fallback that achieves opacity: resorting to a global lock.
Every time a transaction acquires the global lock, all hardware transactions abort and only retry when the lock is released.

In \acrshort{htm}, the commit operation is implemented entirely in hardware.
This poses a challenge on how to implement ordered commits because we introduce conflicts if transactions signal each other whose turn it is to commit using a shared variable.
For example, imagine two non-conflicting transactions~$t_1$~and~$t_2$, serialized in that order.
Transaction~$t_2$~attempts to commit before~$t_1$.
It reads the commit-order variable,~$sn_c$,~and observes that it is still not its turn so it waits, e.g., because~$t_2$~can only commit when~$sn_c = 1$.
When transaction~$t_1$~commits it sets~$sn_c = 1$, triggering a conflict in~$t_2$~because it observed a (now) stale value.
Implementing fast transactions is also challenging because all concurrency control is performed by the hardware.

To implement our prototype we looked at the existing \acrshort{htm}s from Intel~\cite{tsx-sc-2013} and IBM~\cite{tm-powerpc-isca-2013}.
To implement ordered commits without inducing false aborts we require the possibility to perform non-transactional accesses, i.e. that do not trigger transactional conflicts.
Unfortunately, Intel provides no support for non-transactional accesses.
However, IBM's \acrshort{htm} has two instructions,~{\texttt{tsuspend}} and~{\texttt{tresume}}, that allow the possibility to suspend, and resume, transactional execution inside a hardware transaction.
(While in suspended mode accesses are performed non-transactionally.)

IBM's \acrshort{htm} also provides a special kind of transaction called \acrfull{rot}.
According to IBM, \acrshort{rot}s are intended to be used for single thread algorithmic speculation~\cite{tm-powerpc-isca-2013}.
For this reason, \acrshort{rot}s also buffer transactional writes in the cache but do not maintain a read~set.
Furthermore, \acrshort{rot}s do not observe buffered transactional writes from other transactions and all writes performed by a \acrshort{rot} become visible to other transactions atomically, making them a prime choice to implement fast transactions.
However, note that \acrshort{rot}s may nevertheless abort due to write-write conflicts with other concurrent transactions.
For these reasons we implemented our prototype on IBM's \acrshort{htm}.
It is still possible to implement Pot with Intel's \acrshort{htm}, albeit with ordered commits inducing false aborts and fast transactions being regular transactions. 

\noindent{\textbf{Baseline \acrshort{htm} transaction.}} 
The relevant IBM's \acrshort{htm} instructions are~{\texttt{tbegin}} and~{\texttt{tcommit}}, to start and commit a hardware transaction, respectively.
We initialize two important variables, $path_t$ and~$tries_t$, when the application starts a transaction, by invoking our usual~${txn\_start}$ operation.
$path_t$~is either HW or SW depending on whether the transaction will execute as a hardware transaction or by software using the global~lock fallback.
$tries_t$~holds the number of remaining attempts to execute the transaction in hardware until we fallback to software~(Fig.~\ref{subfig:pcc_htm_orig},~lines~2--4).
(We retry 10 times like in GCC's experimental implementation.)
If there is an ongoing transaction executing in software, we wait until the global~lock is free, otherwise the hardware transaction may observe an inconsistent state and violate opacity~(line~5).
After the lock is free we start a hardware transaction by issuing the~{\texttt{tbegin}} instruction~(line~6).
From this point on, every memory access is performed transactionally.
Finally, we subscribe the global~lock by checking if it is locked before proceeding with the actual application code~(lines~7--8).
%
%
%
By checking if the lock is taken it becomes part of the transaction's read~set, so if any transaction falls back to software, any active hardware transaction is immediately aborted.

Committing a transaction depends on whether it executed in hardware~($path_t =$ HW) or software~(SW).
We commit a hardware transaction using the~{\texttt{tcommit}} instruction, whereas for a software transaction we simply release the global~lock~(lines~20--23).
Note that the~{\texttt{tcommit}} operation may still trigger an abort if the transaction fails to commit.

When a hardware transaction aborts, the control flow jumps to the~$txn\_abort$~handler.
First, we check whether the abort is expected to be persistent by inspecting the IBM's TEXASR register, which contains several hints about the reason why the transaction aborted.
For example, an abort due to capacity restrictions is persistent.
If the abort is persistent, we fallback to software by acquiring the global~lock and execute the transaction's application code~(lines~11,~15--18).
Otherwise, we decrement the number of remaining attempts and control flow jumps back to $txn\_start$.

\noindent{\textbf{Speculative \acrshort{htm} transaction.}} 
Like in our \acrshort{stm} implementation, when a transaction starts for the first time it requests a sequence number from the sequencer~(Fig.~\ref{subfig:pcc_htm_spec},~line~5).
After waiting until the global~lock is free, we check whether it is the transaction's turn to commit.
If so, we begin a \acrshort{rot} and switch to fast mode~(lines~8--9).
Otherwise, we sample the sequencer number of the current fast transaction~$sn$~(we explain why shortly), then begin a hardware transaction, and subscribe to the global~lock~(lines~11--13).
To commit a transaction we: (1)~issue the~{\texttt{tsuspend}} instruction to suspend transactional execution~(line~22), (2)~wait for our turn to commit~(line~23), (3)~issue the~{\texttt{tresume}} instruction to resume transactional execution~(line~24),~and (4)~issue the~{\texttt{tcommit}} instruction to commit~(line~25).
If the commit is successful we update the~$sn_c$~variable accordingly~(line~26).

If the speculative transaction aborts due to persistent reasons, there is no point in retrying the transaction until it is it's turn to commit~(line~18).
Otherwise, we wait until the concurrent fast transaction commits to retry the speculative transaction~(line~20)---recall that we sampled its sequence number~$sn$~when we started the aborted hardware transaction~(line~11).
The rationale for waiting for the concurrent fast transaction to commit is to minimize the chances of aborting the fast transaction via write-write conflicts.

\noindent{\textbf{Fast \acrshort{htm} transaction.}} 
As previously stated, fast transactions execute as \acrshort{rot}s.
Unlike regular hardware transactions, \acrshort{rot}s do not maintain a read~set so they enjoy an increased capacity limit that can be used exclusively for writes.
Transactions that previously exceeded capacity constraints and had to fallback to software might now be able to commit in hardware.
This has the potential to increase the parallelism in the system because falling back to software effectively ``stops the world.''

Committing a fast transaction is essentially equivalent to the standard \acrshort{htm} transaction with an additional update to~$sn_c$~(Fig.~\ref{subfig:pcc_htm_fast}~lines~14--18).
If a fast transaction aborts due to capacity restrictions it falls back to software~(lines~5--6,~9--12).

Note that the hardware ensures that the fast transaction's reads do \emph{not} observe the writes of concurrent speculative transactions.
Moreover, if the fast transaction reads a memory location that has been written by a concurrent speculative transaction, the hardware aborts the speculative transaction immediately if it is executing, or when it issues the \texttt{tresume} instruction if it is suspended.
%
\section{Experimental evaluation} 
\label{sec:evaluation}
All experiments were run on a 10-core IBM POWER8 with a total of 128GB RAM.
%
%
We highlight the fact that the machine has a NUMA architecture.
Particularly, the memory latencies in our experiments are as follows: with 1 to 4 threads memory latencies are uniform, while with 8 or more threads memory latencies increase up to~$2 \times$.

We evaluate Pot using the popular STAMP 0.9.10 benchmark suite~\cite{stamp-iiswc-2008} and STMBench7~\cite{stmbench7-eurosys-2007}, using the parameters listed in Fig.~\ref{fig:evaluation_stm_params}.
STAMP consists of 8~representative applications from different domains, e.g. online transaction processing, iterative clustering algorithms, and Delaunay mesh refinement~(Vacation, KMeans, and Yada, resp.)~\cite{stamp-iiswc-2008}.
Some STAMP benchmarks, such as Labyrinth, KMeans, and Yada, output non-deterministic results using STM.
The benefits of Pot in these benchmarks are that the computed Labyrinth's solution, KMeans' clusters, and Yada's mesh, are always the same across executions. 
\begin{figure}[tb]
    \centering
    \footnotesize
\begin{tabular}{rl}
\toprule
    Benchmark      & Parameters \\
\midrule
    Bayes          & \texttt{-v 32 -r 4096 -n 10 -p 40 -i 2 -e 8 -s 1} \\
    Genome         & \texttt{-g 65536 -s 32 -n 16777216} \\
    Intruder       & \texttt{-a 10 -l 2048 -n 8192 -s 1} \\
    Kmeans$-$      & \texttt{-m 40 -n 40 -t 0.00001 -i inputs/random-n65536-d32-c16.txt} \\
    KMeans$+$      & \texttt{-m 15 -n 15 -t 0.00001 -i inputs/random-n65536-d32-c16.txt} \\
    Labyrinth      & \texttt{-i inputs/random-x512-y512-z7-n512.txt} \\
    SSCA2          & \texttt{-s 20 -i 1.0 -u 1.0 -l 3 -p 3} \\
    Vacation$-$    & \texttt{-n 8 -q 90 -u 98 -r 1048576 -t 4194304} \\
    Vacation$+$    & \texttt{-n 8 -q 10 -u 90 -r 1048576 -t 4194304} \\
    Yada           & \texttt{-a 15 -i inputs/ttimeu1000000.2} \\
    STMBench7 (r)  & \texttt{-t true -w r} \\
    STMBench7 (rw) & \texttt{-t true -w rw} \\
    STMBench7 (w)  & \texttt{-t true -w w} \\
\bottomrule
\end{tabular}
    \caption{Parameters used in STAMP and STMBench7.}
    \label{fig:evaluation_stm_params}
\end{figure}
STMBench7 is a more complex benchmark suggestive of CAD, CAM or CASE software~\cite{stmbench7-eurosys-2007}.
Results are the average of five runs.
The GCC version is Red~Hat~5.1.1-4.
\subsection{\Acrlong{stm}} 
\label{sub:evaluation_stm}
In this section we evaluate our Pot \acrshort{stm} prototype.
We seek to answer the following questions:

\noindent{\textbf{Are fast transactions effective?~(\S\ref{subsub:evaluation_stm_fast}.)}}
Yes, they successfully reduce concurrency control overheads and execute faster than regular transactions.
Our experiments show that fast transactions already execute faster than regular transactions even when they perform as little as 1~read and 1~write access, despite the addition work performed regarding the sequencer and switching modes~(Fig.~\ref{fig:evaluation_stm_fast}).

\noindent{\textbf{Does Pot ensure determinism efficiently?~(\S\ref{subsub:evaluation_stm_perf}.)}}
We argue that it does.
Our experiments show that Pot ensures deterministic execution across all of STAMP's benchmarks with an average slowdown over nondeterministic execution of less than~${2 \times}$~(geometric mean of Fig.~\ref{fig:evaluation_perf_stm_stamp}), and it is~${\approx 5 \times}$ \emph{faster} on average than the nondeterministic baseline in STMBench7~(geometric mean of Fig.~\ref{fig:evaluation_perf_stm_sb7}).

\noindent{\textbf{Does Pot improve upon the state of the art?~~(\S\ref{subsub:evaluation_stm_perf}.)}}
Yes, Pot successfully lowers the overheads of ensuring determinism when compared with DeSTM~\cite{destm-pact-2014}.
Our experiments show that, when compared to DeSTM, Pot is up to $\approx 3 \times$ faster than DeSTM on average across the STAMP benchmarks~(geometric mean of Fig.~\ref{fig:evaluation_perf_stm_stamp}) and up to $\approx 9 \times$ faster on average in STMBench7~(geometric mean of Fig.~\ref{fig:evaluation_perf_stm_sb7}), and scales better with the number of threads~(Fig.~\ref{fig:evaluation_scalability_stm_stamp} and~\ref{fig:evaluation_scalability_stm_sb7}).
\subsubsection{Effectiveness of fast transactions} 
\label{subsub:evaluation_stm_fast}
\begin{figure}[tb]
    \centering
    \begin{tikzpicture}
        \begin{axis}[
            xlabel={Number of accesses},
            xmin=0, xmax=7,
             xtick={0,1,2,3,4,5,6,7},
             xticklabels={0,1,2,4,8,16,32,64},
             x tick label style={font=\footnotesize},
             x label style={font=\footnotesize},
             y label style={font=\footnotesize,align=center},
             ylabel near ticks,
             ylabel={$\times$ faster than \\ \acrshort{stm} transaction},
             ylabel style={font=\footnotesize,align=center},
             ymin=0, ymax=11,
             ytick={0,1,2,3,4,5,6,7,8,9,10,11},
             y tick label style={font=\footnotesize},
             ymajorgrids=true,
             legend columns=1,
             legend style={font=\footnotesize,fill=none,draw=none},
             legend pos=north west,
             title style={font=\footnotesize},
             grid style={dotted,gray!50},
             width=0.63\textwidth,
             height=0.33\textwidth
         ]
         \addlegendentry{0\% reads}
         \addplot[color=darkcyan,thick,mark size=3pt,mark=triangle] coordinates {
             (0,.97  )   
             (1,1.96 )   
             (2,2.53 )   
             (3,3.43 )   
             (4,4.25 )   
             (5,6.01 )   
             (6,7.90 )   
             (7,10.36)   
         };
         \addlegendentry{50\% reads}
         \addplot[color=darkgoldenrod,thick,mark=square] coordinates {
             (0,1.07)   
             (1,1.10)   
             (2,2.12)   
             (3,2.59)   
             (4,3.20)   
             (5,3.30)   
             (6,4.16)   
             (7,5.12)   
         };
         \addlegendentry{100\% reads}
         \addplot[color=darkcoral,thick,mark size=3pt,mark=x] coordinates {
             (0,.90 )   
             (1,.98 )   
             (2,1.06)   
             (3,1.43)   
             (4,1.32)   
             (5,1.42)   
             (6,1.48)   
             (7,1.52)   
         };
         \end{axis}
     \end{tikzpicture}
     \caption{Speedup achieved by a Pot fast transaction over the baseline \acrshort{stm} transaction.}
     \label{fig:evaluation_stm_fast}
\end{figure}
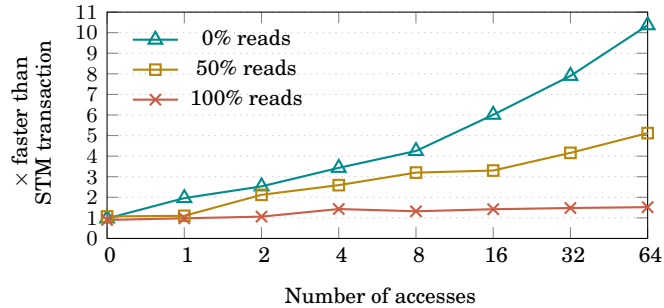
The fast transaction's objective is to reduce concurrency control overheads in order to mitigate the potential loss of parallelism introduced by ordered commits.
To measure how effective is the fast execution mode we executed a microbenchmark that consists of a simple key-value data structure implemented with an array of counters.
We use a single thread, and vary the number of accesses performed by transactions, and the accesses' read/write~ratio.

Fig.~\ref{fig:evaluation_stm_fast} shows how much faster the Pot fast transaction protocol is than the baseline \acrshort{stm} transaction.
Transactions with~0~accesses consist of~$txn\_begin$ immediately followed by~$txn\_commit$.
This allows us to measure the overhead imposed by the additional work performed by the sequencer, ordered~commits, and transaction~modes, which is negligible.
By increasing the number of accesses we observe that, as expected, fast transactions perform increasingly better than the baseline.
We also observe that write operations contribute more to the achieved speedup.
This is due to the fact that in the baseline STM write operations impose overhead on reads because reads must query the write~set for possible buffered values.
Write operations also impose overhead on the commit operation due to the need to lock the write~set, perform the write~back, and unlock the write~set.
Fast transactions bypass all these sources of overhead.
However, fast transactions do not achieve observable gains when transactions are read-only.
This is because read-only transactions in the baseline STM do not need to validate the read~set at commit time---they are serialized at begin time.
Overall, fast transactions are successful in minimizing concurrency control overheads, even for transactions that perform as little as 1~read and 1~write.
\subsubsection{Comparison with the state of the art} 
\label{subsub:evaluation_stm_perf}
In this section we evaluate deterministic execution using Pot in the popular STAMP 0.9.10 benchmark suite~\cite{stamp-iiswc-2008}, and STMBench7~\cite{stmbench7-eurosys-2007}.
%
%
%
%
We also compare Pot against DeSTM, a state of the art system in deterministic execution of \acrshort{stm} programs.
To perform an apples-to-apples comparison, we implemented DeSTM in our own prototype.\footnote{\small DeSTM is not publicly available. We asked the authors for the source code via e-mail but got no response.}
Both Pot and DeSTM are based on the same baseline STM protocol and use the exact same sequencer.
See \S\ref{sec:related} for a comparison of Pot with DeSTM.
We also implemented a deterministic and non-speculative solution based on a global~lock that transactions acquire according to the order defined by the sequencer, i.e. transactions acquire a global~lock at $txn\_begin$ and release it at $txn\_commit$~(PoGL, as in Preordered Global Lock).
The rationale is that PoGL is a ``trivial'' implementation of PCC without any speculation.
We show results for DeSTM, PoGL, ordered commits only~(Pot$-$), ordered commits and transaction modes~(Pot$*$), and ordered commits, transaction modes, and live promotion~(Pot).

\noindent{\textbf{Performance.}} 
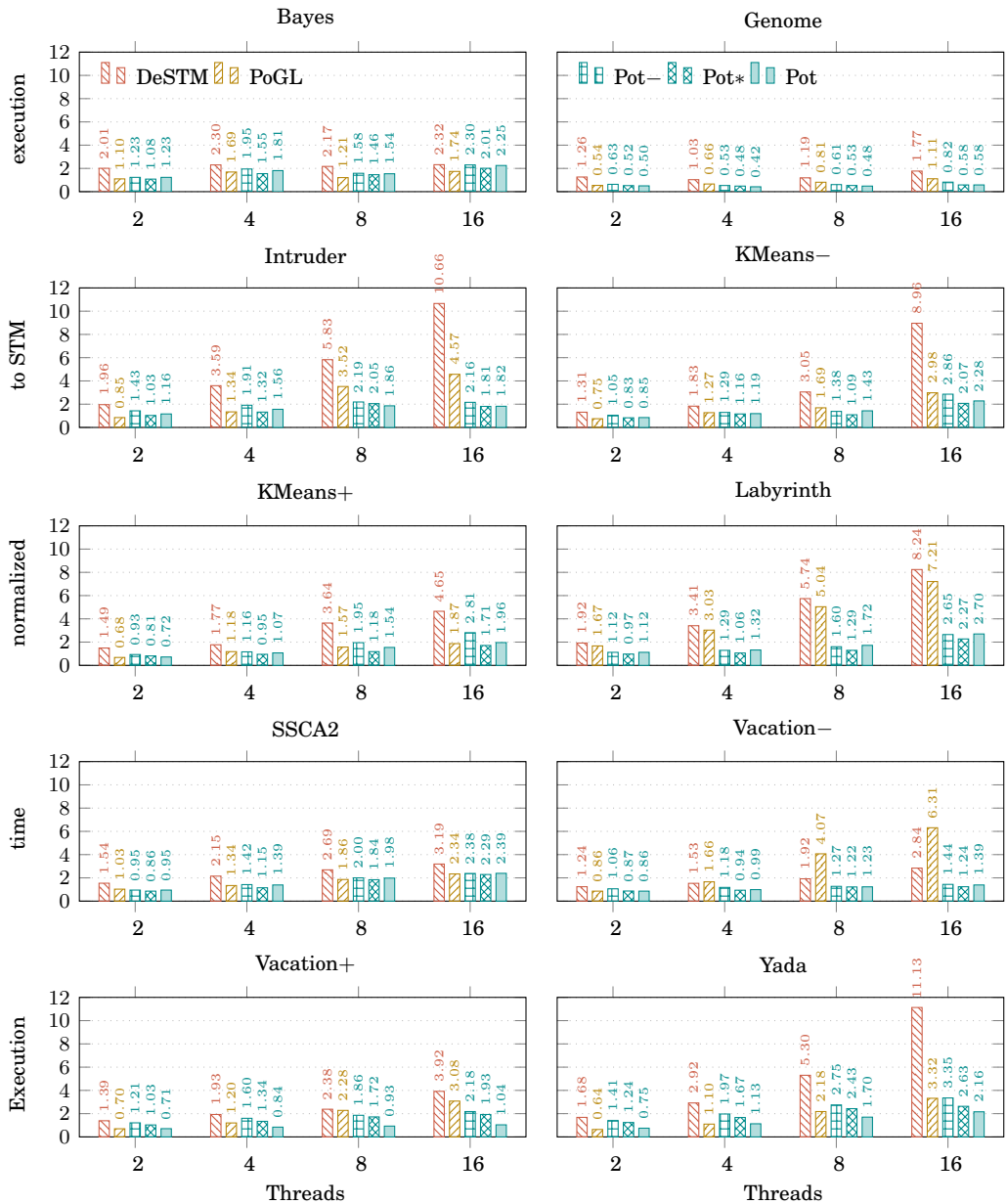
\begin{figure*}[tb]
    \centering
    \begin{tikzpicture}
        \begin{axis}[
            xmin=0.5, xmax=4.5,
            xtick={1,2,3,4},
            xticklabels={2,4,8,16},
            x tick label style={font=\footnotesize},
            x label style={font=\footnotesize},
            ymin=0, ymax=12,
            ytick={0,2,4,6,8,10,12},
            ymajorgrids=true,
            y tick label style={font=\footnotesize},
            y label style={align=left,font=\footnotesize},
            ylabel near ticks,
            ylabel={execution},
            ylabel style={align=center,font=\footnotesize},
            legend columns=2,
            legend style={font=\footnotesize,fill=none,draw=none},
            legend pos=north west,
            title style={font=\footnotesize},
            grid style={dotted,gray!50},
            width=0.55\textwidth,
            height=0.25\textwidth,
            ybar,
            bar width=4pt,
            nodes near coords,
            every node near coord/.append style={font=\tiny,/pgf/number format/fixed, /pgf/number format/zerofill, /pgf/number format/precision=2, rotate=90, anchor=west},
            title={Bayes},
        ]
            \addlegendentry{DeSTM}
            \addplot[color=darkcoral,pattern=north west lines,pattern color=darkcoral] coordinates {
                (1, 2.01) 
                (2, 2.30) 
                (3, 2.17) 
                (4, 2.32) 
                
            };
			\addlegendentry{PoGL}
            \addplot[color=darkgoldenrod,pattern=north east lines,pattern color=darkgoldenrod] coordinates {
                (1, 1.10) 
                (2, 1.69) 
                (3, 1.21) 
                (4, 1.74) 
                
            };
            \addlegendentry{Pot$-$}
            \addplot[color=darkcyan,pattern=grid,pattern color=darkcyan] coordinates {
                (1, 1.23) 
                (2, 1.95) 
                (3, 1.58) 
                (4, 2.30) 
            };
            \addlegendentry{Pot$*$}
            \addplot[color=darkcyan,pattern=crosshatch,pattern color=darkcyan] coordinates {
                (1, 1.08) 
                (2, 1.55) 
                (3, 1.46) 
                (4, 2.01) 
            };
            \addlegendentry{Pot}
            \addplot[color=darkcyan,fill=darkcyan!30] coordinates {
                (1, 1.23) 
                (2, 1.81) 
                (3, 1.54) 
                (4, 2.25) 
            };
            \legend{DeSTM,PoGL,,,}
        \end{axis}
    \end{tikzpicture}
    \begin{tikzpicture}
        \begin{axis}[
            xmin=0.5, xmax=4.5,
            xtick={1,2,3,4},
            xticklabels={2,4,8,16},
            x tick label style={font=\footnotesize},
            x label style={font=\footnotesize},
            ymin=0, ymax=12,
            ytick={0,2,4,6,8,10,12},
            yticklabels={,,,,,},
            ymajorgrids=true,
            y tick label style={font=\footnotesize},
            y label style={align=left,font=\footnotesize},
            ylabel near ticks,
            ylabel style={align=center,font=\footnotesize},
            legend columns=3,
            legend style={font=\footnotesize,fill=none,draw=none},
            legend pos=north west,
            title style={font=\footnotesize},
            grid style={dotted,gray!50},
            width=0.55\textwidth,
            height=0.25\textwidth,
            ybar,
            bar width=4pt,
            nodes near coords,
            every node near coord/.append style={font=\tiny,/pgf/number format/fixed, /pgf/number format/zerofill, /pgf/number format/precision=2, rotate=90, anchor=west},
            title={Genome},
        ]
            \addlegendentry{DeSTM}
            \addplot[color=darkcoral,pattern=north west lines,pattern color=darkcoral] coordinates {
                (1, 1.26) 
                (2, 1.03) 
                (3, 1.19) 
                (4, 1.77) 
            };
			\addlegendentry{PoGL}
            \addplot[color=darkgoldenrod,pattern=north east lines,pattern color=darkgoldenrod] coordinates {
                (1,  .54) 
                (2,  .66) 
                (3,  .81) 
                (4, 1.11) 
                
            };
            \addlegendentry{Pot$-$}
            \addplot[color=darkcyan,pattern=grid,pattern color=darkcyan] coordinates {
                (1, .63) 
                (2, .53) 
                (3, .61) 
                (4, .82) 
            };
            \addlegendentry{Pot$*$}
            \addplot[color=darkcyan,pattern=crosshatch,pattern color=darkcyan] coordinates {
                (1, .52) 
                (2, .48) 
                (3, .53) 
                (4, .58) 
            };
            \addlegendentry{Pot}
            \addplot[color=darkcyan,fill=darkcyan!30] coordinates {
                (1, .50) 
                (2, .42) 
                (3, .48) 
                (4, .58) 
            };
            \legend{,,Pot$-$,Pot$*$,Pot}
        \end{axis}
    \end{tikzpicture}
    \begin{tikzpicture}
        \begin{axis}[
            xmin=0.5, xmax=4.5,
            xtick={1,2,3,4},
            xticklabels={2,4,8,16},
            x tick label style={font=\footnotesize},
            x label style={font=\footnotesize},
            ymin=0, ymax=12,
            ytick={0,2,4,6,8,10,12},
            ymajorgrids=true,
            y tick label style={font=\footnotesize},
            y label style={align=left,font=\footnotesize},
            ylabel near ticks,
            ylabel={to STM},
            ylabel style={align=center,font=\footnotesize},
            legend style={font=\footnotesize,fill=none,draw=none},
            legend pos=north west,
            title style={font=\footnotesize},
            grid style={dotted,gray!50},
            width=0.55\textwidth,
            height=0.25\textwidth,
            ybar,
            bar width=4pt,
            nodes near coords,
            every node near coord/.append style={font=\tiny,/pgf/number format/fixed, /pgf/number format/zerofill, /pgf/number format/precision=2, rotate=90, anchor=west},
            title={Intruder},
        ]
            \addlegendentry{DeSTM}
            \addplot[color=darkcoral,pattern=north west lines,pattern color=darkcoral] coordinates {
                (1, 1.96 ) 
                (2, 3.59 ) 
                (3, 5.83 ) 
                (4, 10.66) 
            };
			\addlegendentry{PoGL}
            \addplot[color=darkgoldenrod,pattern=north east lines,pattern color=darkgoldenrod] coordinates {
                (1,  .85) 
                (2, 1.34) 
                (3, 3.52) 
                (4, 4.57) 
                
            };
            \addlegendentry{Pot$-$}
            \addplot[color=darkcyan,pattern=grid,pattern color=darkcyan] coordinates {
                (1, 1.43) 
                (2, 1.91) 
                (3, 2.19) 
                (4, 2.16) 
            };
            \addlegendentry{Pot$*$}
            \addplot[color=darkcyan,pattern=crosshatch,pattern color=darkcyan] coordinates {
                (1, 1.03) 
                (2, 1.32) 
                (3, 2.05) 
                (4, 1.81) 
            };
            \addlegendentry{Pot}
            \addplot[color=darkcyan,fill=darkcyan!30] coordinates {
                (1, 1.16) 
                (2, 1.56) 
                (3, 1.86) 
                (4, 1.82) 
            };
            \legend{}
        \end{axis}
    \end{tikzpicture}
    \begin{tikzpicture}
        \begin{axis}[
            xmin=0.5, xmax=4.5,
            xtick={1,2,3,4},
            xticklabels={2,4,8,16},
            x tick label style={font=\footnotesize},
            x label style={font=\footnotesize},
            ymin=0, ymax=12,
            ytick={0,2,4,6,8,10,12},
            yticklabels={,,,,,},
            ymajorgrids=true,
            y tick label style={font=\footnotesize},
            y label style={align=left,font=\footnotesize},
            ylabel near ticks,
            ylabel style={align=center,font=\footnotesize},
            legend style={font=\footnotesize,fill=none,draw=none},
            legend pos=north west,
            title style={font=\footnotesize},
            grid style={dotted,gray!50},
            width=0.55\textwidth,
            height=0.25\textwidth,
            ybar,
            bar width=4pt,
            nodes near coords,
            every node near coord/.append style={font=\tiny,/pgf/number format/fixed, /pgf/number format/zerofill, /pgf/number format/precision=2, rotate=90, anchor=west},
            title={KMeans$-$},
        ]
            \addlegendentry{DeSTM}
            \addplot[color=darkcoral,pattern=north west lines,pattern color=darkcoral] coordinates {
                (1, 1.31) 
                (2, 1.83) 
                (3, 3.05) 
                (4, 8.96) 
            };
			\addlegendentry{PoGL}
            \addplot[color=darkgoldenrod,pattern=north east lines,pattern color=darkgoldenrod] coordinates {
                (1,  .75) 
                (2, 1.27) 
                (3, 1.69) 
                (4, 2.98) 
                
            };
            \addlegendentry{Pot$-$}
            \addplot[color=darkcyan,pattern=grid,pattern color=darkcyan] coordinates {
                (1, 1.05) 
                (2, 1.29) 
                (3, 1.38) 
                (4, 2.86) 
            };
            \addlegendentry{Pot$*$}
            \addplot[color=darkcyan,pattern=crosshatch,pattern color=darkcyan] coordinates {
                (1, .83 ) 
                (2, 1.16) 
                (3, 1.09) 
                (4, 2.07) 
            };
            \addlegendentry{Pot}
            \addplot[color=darkcyan,fill=darkcyan!30] coordinates {
                 (1, .85 ) 
                 (2, 1.19) 
                 (3, 1.43) 
                 (4, 2.28) 
            };
            \legend{}
        \end{axis}
    \end{tikzpicture}
    \begin{tikzpicture}
        \begin{axis}[
            xmin=0.5, xmax=4.5,
            xtick={1,2,3,4},
            xticklabels={2,4,8,16},
            x tick label style={font=\footnotesize},
            x label style={font=\footnotesize},
            ymin=0, ymax=12,
            ytick={0,2,4,6,8,10,12},
            ymajorgrids=true,
            y tick label style={font=\footnotesize},
            y label style={align=left,font=\footnotesize},
            ylabel near ticks,
            ylabel={normalized},
            ylabel style={align=center,font=\footnotesize},
            legend style={font=\footnotesize,fill=none,draw=none},
            legend pos=north west,
            title style={font=\footnotesize},
            grid style={dotted,gray!50},
            width=0.55\textwidth,
            height=0.25\textwidth,
            ybar,
            bar width=4pt,
            nodes near coords,
            every node near coord/.append style={font=\tiny,/pgf/number format/fixed, /pgf/number format/zerofill, /pgf/number format/precision=2, rotate=90, anchor=west},
            title={KMeans$+$},
        ]
            \addlegendentry{DeSTM}
            \addplot[color=darkcoral,pattern=north west lines,pattern color=darkcoral] coordinates {
                (1, 1.49) 
                (2, 1.77) 
                (3, 3.64) 
                (4, 4.65) 
            };
			\addlegendentry{PoGL}
            \addplot[color=darkgoldenrod,pattern=north east lines,pattern color=darkgoldenrod] coordinates {
                (1,  .68) 
                (2, 1.18) 
                (3, 1.57) 
                (4, 1.87) 
                
            };
            \addlegendentry{Pot$-$}
            \addplot[color=darkcyan,pattern=grid,pattern color=darkcyan] coordinates {
                (1, .93 ) 
                (2, 1.16) 
                (3, 1.95) 
                (4, 2.81) 
            };
            \addlegendentry{Pot$*$}
            \addplot[color=darkcyan,pattern=crosshatch,pattern color=darkcyan] coordinates {
                (1, .81 ) 
                (2, .95 ) 
                (3, 1.18) 
                (4, 1.71) 
            };
            \addlegendentry{Pot}
            \addplot[color=darkcyan,fill=darkcyan!30] coordinates {
                 (1, .72 ) 
                 (2, 1.07) 
                 (3, 1.54) 
                 (4, 1.96) 
            };
            \legend{}
        \end{axis}
    \end{tikzpicture}
    \begin{tikzpicture}
        \begin{axis}[
            xmin=0.5, xmax=4.5,
            xtick={1,2,3,4},
            xticklabels={2,4,8,16},
            x tick label style={font=\footnotesize},
            x label style={font=\footnotesize},
            ymin=0, ymax=12,
            ytick={0,2,4,6,8,10,12},
            yticklabels={,,,,,},
            ymajorgrids=true,
            y tick label style={font=\footnotesize},
            y label style={align=left,font=\footnotesize},
            ylabel near ticks,
            ylabel style={align=center,font=\footnotesize},
            legend style={font=\footnotesize,fill=none,draw=none},
            legend pos=north west,
            title style={font=\footnotesize},
            grid style={dotted,gray!50},
            width=0.55\textwidth,
            height=0.25\textwidth,
            ybar,
            bar width=4pt,
            nodes near coords,
            every node near coord/.append style={font=\tiny,/pgf/number format/fixed, /pgf/number format/zerofill, /pgf/number format/precision=2, rotate=90, anchor=west},
            title={Labyrinth},
        ]
            \addlegendentry{DeSTM}
            \addplot[color=darkcoral,pattern=north west lines,pattern color=darkcoral] coordinates {
                (1, 1.92) 
                (2, 3.41) 
                (3, 5.74) 
                (4, 8.24) 
            };
			\addlegendentry{PoGL}
            \addplot[color=darkgoldenrod,pattern=north east lines,pattern color=darkgoldenrod] coordinates {
                (1, 1.67) 
                (2, 3.03) 
                (3, 5.04) 
                (4, 7.21) 
                
            };
            \addlegendentry{Pot$-$}
            \addplot[color=darkcyan,pattern=grid,pattern color=darkcyan] coordinates {
                (1, 1.12) 
                (2, 1.29) 
                (3, 1.60) 
                (4, 2.65) 
            };
            \addlegendentry{Pot$*$}
            \addplot[color=darkcyan,pattern=crosshatch,pattern color=darkcyan] coordinates {
                (1, .97 ) 
                (2, 1.06) 
                (3, 1.29) 
                (4, 2.27) 
            };
            \addlegendentry{Pot}
            \addplot[color=darkcyan,fill=darkcyan!30] coordinates {
                (1, 1.12) 
                (2, 1.32) 
                (3, 1.72) 
                (4, 2.70) 
            };
            \legend{}
        \end{axis}
    \end{tikzpicture}
    \begin{tikzpicture}
        \begin{axis}[
            xmin=0.5, xmax=4.5,
            xtick={1,2,3,4},
            xticklabels={2,4,8,16},
            x tick label style={font=\footnotesize},
            x label style={font=\footnotesize},
            ymin=0, ymax=12,
            ytick={0,2,4,6,8,10,12},
            ymajorgrids=true,
            y tick label style={font=\footnotesize},
            y label style={align=left,font=\footnotesize},
            ylabel near ticks,
            ylabel={time},
            ylabel style={align=center,font=\footnotesize},                        
            legend style={font=\footnotesize,fill=none,draw=none},
            legend pos=north west,
            title style={font=\footnotesize},
            grid style={dotted,gray!50},
            width=0.55\textwidth,
            height=0.25\textwidth,
            ybar,
            bar width=4pt,
            nodes near coords,
            every node near coord/.append style={font=\tiny,/pgf/number format/fixed, /pgf/number format/zerofill, /pgf/number format/precision=2, rotate=90, anchor=west},
            title={SSCA2},
        ]
            \addlegendentry{DeSTM}
            \addplot[color=darkcoral,pattern=north west lines,pattern color=darkcoral] coordinates {
                (1, 1.54) 
                (2, 2.15) 
                (3, 2.69) 
                (4, 3.19) 
            };
			\addlegendentry{PoGL}
            \addplot[color=darkgoldenrod,pattern=north east lines,pattern color=darkgoldenrod] coordinates {
                (1, 1.03) 
                (2, 1.34) 
                (3, 1.86) 
                (4, 2.34) 
                
            };
            \addlegendentry{Pot$-$}
            \addplot[color=darkcyan,pattern=grid,pattern color=darkcyan] coordinates {
                (1, .95 ) 
                (2, 1.42) 
                (3, 2.00) 
                (4, 2.38) 
            };
            \addlegendentry{Pot$*$}
            \addplot[color=darkcyan,pattern=crosshatch,pattern color=darkcyan] coordinates {
                (1, .86 ) 
                (2, 1.15) 
                (3, 1.84) 
                (4, 2.29) 
            };
            \addlegendentry{Pot}
            \addplot[color=darkcyan,fill=darkcyan!30] coordinates {
                (1, .95 ) 
                (2, 1.39) 
                (3, 1.98) 
                (4, 2.39) 
            };
            \legend{}
        \end{axis}
    \end{tikzpicture}
    \begin{tikzpicture}
        \begin{axis}[
            xmin=0.5, xmax=4.5,
            xtick={1,2,3,4},
            xticklabels={2,4,8,16},
            x tick label style={font=\footnotesize},
            x label style={font=\footnotesize},
            ymin=0, ymax=12,
            ytick={0,2,4,6,8,10,12},
            yticklabels={,,,,,},
            ymajorgrids=true,
            y tick label style={font=\footnotesize},
            y label style={align=left,font=\footnotesize},
            ylabel near ticks,
            ylabel style={align=center,font=\footnotesize},
            legend style={font=\footnotesize,fill=none,draw=none},
            legend pos=north west,
            title style={font=\footnotesize},
            grid style={dotted,gray!50},
            width=0.55\textwidth,
            height=0.25\textwidth,
            ybar,
            bar width=4pt,
            nodes near coords,
            every node near coord/.append style={font=\tiny,/pgf/number format/fixed, /pgf/number format/zerofill, /pgf/number format/precision=2, rotate=90, anchor=west},
            title={Vacation$-$},
        ]
            \addlegendentry{DeSTM}
            \addplot[color=darkcoral,pattern=north west lines,pattern color=darkcoral] coordinates {
                (1, 1.24) 
                (2, 1.53) 
                (3, 1.92) 
                (4, 2.84) 
            };
			\addlegendentry{PoGL}
            \addplot[color=darkgoldenrod,pattern=north east lines,pattern color=darkgoldenrod] coordinates {
                (1,  .86) 
                (2, 1.66) 
                (3, 4.07) 
                (4, 6.31) 
                
            };
            \addlegendentry{Pot$-$}
            \addplot[color=darkcyan,pattern=grid,pattern color=darkcyan] coordinates {
                (1, 1.06) 
                (2, 1.18) 
                (3, 1.27) 
                (4, 1.44) 
            };
            \addlegendentry{Pot$*$}
            \addplot[color=darkcyan,pattern=crosshatch,pattern color=darkcyan] coordinates {
                (1, .87 ) 
                (2, .94 ) 
                (3, 1.22) 
                (4, 1.24) 
            };
            \addlegendentry{Pot}
            \addplot[color=darkcyan,fill=darkcyan!30] coordinates {
                (1, .86 ) 
                (2, .99 ) 
                (3, 1.23) 
                (4, 1.39) 
            };
            \legend{}
        \end{axis}
    \end{tikzpicture}
    \begin{tikzpicture}
        \begin{axis}[
            xmin=0.5, xmax=4.5,
            xtick={1,2,3,4},
            xticklabels={2,4,8,16},
            x tick label style={font=\footnotesize},
            x label style={font=\footnotesize},
            xlabel={Threads},
            ymin=0, ymax=12,
            ytick={0,2,4,6,8,10,12},
            ymajorgrids=true,
            y tick label style={font=\footnotesize},
            y label style={align=left,font=\footnotesize},
            ylabel near ticks,
            ylabel={Execution},
            ylabel style={align=center,font=\footnotesize},
            legend style={font=\footnotesize,fill=none,draw=none},
            legend pos=north west,
            title style={font=\footnotesize},
            grid style={dotted,gray!50},
            width=0.55\textwidth,
            height=0.25\textwidth,
            ybar,
            bar width=4pt,
            nodes near coords,
            every node near coord/.append style={font=\tiny,/pgf/number format/fixed, /pgf/number format/zerofill, /pgf/number format/precision=2, rotate=90, anchor=west},
            title={Vacation$+$},
        ]
            \addlegendentry{DeSTM}
            \addplot[color=darkcoral,pattern=north west lines,pattern color=darkcoral] coordinates {
                (1, 1.39) 
                (2, 1.93) 
                (3, 2.38) 
                (4, 3.92) 
            };
			\addlegendentry{PoGL}
            \addplot[color=darkgoldenrod,pattern=north east lines,pattern color=darkgoldenrod] coordinates {
                (1,  .70) 
                (2, 1.20) 
                (3, 2.28) 
                (4, 3.08) 
                
            };
            \addlegendentry{Pot$-$}
            \addplot[color=darkcyan,pattern=grid,pattern color=darkcyan] coordinates {
                (1, 1.21) 
                (2, 1.60) 
                (3, 1.86) 
                (4, 2.18) 
            };
            \addlegendentry{Pot$*$}
            \addplot[color=darkcyan,pattern=crosshatch,pattern color=darkcyan] coordinates {
                (1, 1.03) 
                (2, 1.34) 
                (3, 1.72) 
                (4, 1.93) 
            };
            \addlegendentry{Pot}
            \addplot[color=darkcyan,fill=darkcyan!30] coordinates {
                (1, .71 ) 
                (2, .84 ) 
                (3, .93 ) 
                (4, 1.04) 
            };
            \legend{}
        \end{axis}
    \end{tikzpicture}
    \begin{tikzpicture}
        \begin{axis}[
            xmin=0.5, xmax=4.5,
            xtick={1,2,3,4},
            xticklabels={2,4,8,16},
            x tick label style={font=\footnotesize},
            x label style={font=\footnotesize},
            xlabel={Threads},
            ymin=0, ymax=12,
            ytick={0,2,4,6,8,10,12},
			yticklabels={,,,,,,},
            ymajorgrids=true,
            y tick label style={font=\footnotesize},
            y label style={align=left,font=\footnotesize},
            ylabel near ticks,
            ylabel style={align=center,font=\footnotesize},  
            legend style={font=\footnotesize,fill=none,draw=none},
            legend pos=north west,
            title style={font=\footnotesize},
            grid style={dotted,gray!50},
            width=0.55\textwidth,
            height=0.25\textwidth,
            ybar,
            bar width=4pt,
            nodes near coords,
            every node near coord/.append style={font=\tiny,/pgf/number format/fixed, /pgf/number format/zerofill, /pgf/number format/precision=2, rotate=90, anchor=west},
            title={Yada},
        ]
            \addlegendentry{DeSTM}
            \addplot[color=darkcoral,pattern=north west lines,pattern color=darkcoral] coordinates {
                (1, 1.68 ) 
                (2, 2.92 ) 
                (3, 5.30 ) 
                (4, 11.13) 
            };
			\addlegendentry{PoGL}
            \addplot[color=darkgoldenrod,pattern=north east lines,pattern color=darkgoldenrod] coordinates {
                (1,  .64) 
                (2, 1.10) 
                (3, 2.18) 
                (4, 3.32) 
                
            };
            \addlegendentry{Pot$-$}
            \addplot[color=darkcyan,pattern=grid,pattern color=darkcyan] coordinates {
                (1, 1.41) 
                (2, 1.97) 
                (3, 2.75) 
                (4, 3.35) 
            };
            \addlegendentry{Pot$*$}
            \addplot[color=darkcyan,pattern=crosshatch,pattern color=darkcyan] coordinates {
                (1, 1.24) 
                (2, 1.67) 
                (3, 2.43) 
                (4, 2.63) 
            };
            \addlegendentry{Pot}
            \addplot[color=darkcyan,fill=darkcyan!30] coordinates {
                (1, .75 ) 
                (2, 1.13) 
                (3, 1.70) 
                (4, 2.16) 
            };
            \legend{}
        \end{axis}
    \end{tikzpicture}
    \caption{
	\textbf{How much slower is the execution of each STAMP benchmark with $x$~threads if we want determinism?}
    The y axis measures the execution time using DeSTM, PoGL~(preordered global lock), Pot$-$~(ordered commits), Pot$*$~(ordered commits and transaction modes), and Pot~(ordered commits, transaction modes, and live promotion), normalized to the nondeterministic execution using the baseline STM~(lower is better).
    $-$ and $+$ refer to the relative levels of contention in the configuration.
    }
    \label{fig:evaluation_perf_stm_stamp}
\end{figure*}
\begin{figure*}[tb]
    \begin{tikzpicture}
        \begin{axis}[
            xmin=0.5, xmax=4.5,
            xtick={1,2,3,4},
            xticklabels={,,,},
            x tick label style={font=\footnotesize},
            x label style={font=\footnotesize},
            ymin=0, ymax=20,
            ytick={0,4,8,12,16,20},
            ymajorgrids=true,
            y tick label style={font=\footnotesize},
            y label style={align=left,font=\footnotesize},
            ylabel near ticks,
            ylabel={to STM execution},
            ylabel style={align=center,font=\footnotesize},
            legend columns=2,
            legend style={font=\footnotesize,fill=none,draw=none},
            legend pos=north east,
            title={Read-dominated},
            title style={font=\footnotesize},
            grid style={dotted,gray!50},
            width=0.55\textwidth,
            height=0.25\textwidth,
            ybar,
            bar width=4pt,
            nodes near coords,
            every node near coord/.append style={font=\tiny,/pgf/number format/fixed, /pgf/number format/zerofill, /pgf/number format/precision=2, rotate=90, anchor=west}
        ]
            \addlegendentry{DeSTM}
            \addplot[color=darkcoral,pattern=north west lines,pattern color=darkcoral] coordinates {
                (1, .46 ) 
                (2, 1.14) 
                (3, 1.20) 
                (4, .86 ) 
                
            };
			\addlegendentry{PoGL}
            \addplot[color=darkgoldenrod,pattern=north east lines,pattern color=darkgoldenrod] coordinates {
                (1, 1.58) 
                (2, 3.84) 
                (3, 3.08) 
                (4, 3.92) 
                
            };
            \addlegendentry{Pot$-$}
            \addplot[color=darkcyan,pattern=grid,pattern color=darkcyan] coordinates {
                (1, .45 ) 
                (2, 1.18) 
                (3, 1.15) 
                (4, 1.00) 
            };
            \addlegendentry{Pot$*$}
            \addplot[color=darkcyan,pattern=crosshatch,pattern color=darkcyan] coordinates {
                (1, .59 ) 
                (2, 1.28) 
                (3, 1.71) 
                (4, 1.62) 
            };
            \addlegendentry{Pot}
            \addplot[color=darkcyan,fill=darkcyan!30] coordinates {
                (1, 1.46) 
                (2, 3.69) 
                (3, 3.11) 
                (4, 4.28) 
            };
            \legend{DeSTM,PoGL,,,}
        \end{axis}
    \end{tikzpicture}
    \begin{tikzpicture}
        \begin{axis}[
            xmin=0.5, xmax=4.5,
            xtick={1,2,3,4},
            xticklabels={,,,},
            x tick label style={font=\footnotesize},
            x label style={font=\footnotesize},
            ymin=0, ymax=20,
            ytick={0,4,8,12,16,20},
            yticklabels={,,,,,},
            ymajorgrids=true,
            y tick label style={font=\footnotesize},
            y label style={align=center,font=\footnotesize},
            ylabel near ticks,
            ylabel style={align=center,font=\footnotesize},
            legend columns=3,
            legend style={font=\footnotesize,fill=none,draw=none},
            legend pos=north east,
            title={Read-write},
            title style={font=\footnotesize},
            grid style={dotted,gray!50},
            width=0.55\textwidth,
            height=0.25\textwidth,
            ybar,
            bar width=4pt,
            nodes near coords,
            every node near coord/.append style={font=\tiny,/pgf/number format/fixed, /pgf/number format/zerofill, /pgf/number format/precision=2, rotate=90, anchor=west}
        ]
            \addlegendentry{DeSTM}
            \addplot[color=darkcoral,pattern=north west lines,pattern color=darkcoral] coordinates {
                (1, 1.00) 
                (2, 1.20) 
                (3, .99 ) 
                (4, .44 ) 
                
            };
			\addlegendentry{PoGL}
            \addplot[color=darkgoldenrod,pattern=north east lines,pattern color=darkgoldenrod] coordinates {
                (1, 9.32) 
                (2, 8.38) 
                (3, 7.64) 
                (4, 4.21) 
                
            };
            \addlegendentry{Pot$-$}
            \addplot[color=darkcyan,pattern=grid,pattern color=darkcyan] coordinates {
                (1,  1.74) 
                (2,  1.20) 
                (3, .80  ) 
                (4, .60  ) 
            };
            \addlegendentry{Pot$*$}
            \addplot[color=darkcyan,pattern=crosshatch,pattern color=darkcyan] coordinates {
                (1, 1.59) 
                (2, 1.15) 
                (3, 1.68) 
                (4, 1.03) 
            };
            \addlegendentry{Pot}
            \addplot[color=darkcyan,fill=darkcyan!30] coordinates {
                (1, 8.36) 
                (2, 6.66) 
                (3, 6.65) 
                (4, 4.22) 
            };
            \legend{,,Pot$-$,Pot$*$,Pot}
        \end{axis}
    \end{tikzpicture}
    \begin{tikzpicture}
        \begin{axis}[
            xmin=0.5, xmax=4.5,
            xtick={1,2,3,4},
            xticklabels={2,4,8,16},
            x tick label style={font=\footnotesize},
            x label style={font=\footnotesize},
            xlabel={Threads},
            ymin=0, ymax=20,
            ytick={0,4,8,12,16,20},
            ymajorgrids=true,
            y tick label style={font=\footnotesize},
            y label style={align=left,font=\footnotesize},
            ylabel near ticks,
            ylabel={Throughput normalized},
            ylabel style={align=center,font=\footnotesize},
            legend style={font=\footnotesize,fill=none,draw=none},
            legend pos=north east,
            title={Write-dominated},
            title style={font=\footnotesize},
            grid style={dotted,gray!50},
            width=0.55\textwidth,
            height=0.25\textwidth,
            ybar,
            bar width=4pt,
            nodes near coords,
        ]
            \addlegendentry{DeSTM}
            \addplot[color=darkcoral,pattern=north west lines,pattern color=darkcoral,node near coord style={font=\tiny,/pgf/number format/fixed, /pgf/number format/zerofill, /pgf/number format/precision=2, rotate=90, anchor=west}] coordinates {
                (1, 1.18) 
                (2, 1.00) 
                (3, .80 ) 
                (4, .39 ) 
                
            };
			\addlegendentry{PoGL}
            \addplot[color=darkgoldenrod,pattern=north east lines,pattern color=darkgoldenrod,node near coord style={font=\tiny,/pgf/number format/fixed, /pgf/number format/zerofill, /pgf/number format/precision=2}] coordinates {
                (1, 19.81) 
                (2, 14.92) 
                (3, 9.27) 
                (4, 6.92) 
                
            };
            \addlegendentry{Pot$-$}
            \addplot[color=darkcyan,pattern=grid,pattern color=darkcyan,node near coord style={font=\tiny,/pgf/number format/fixed, /pgf/number format/zerofill, /pgf/number format/precision=2, rotate=90, anchor=west}] coordinates {
                (1, 1.61) 
                (2, 1.11) 
                (3, .81 ) 
                (4, .48 ) 
            };
            \addlegendentry{Pot$*$}
            \addplot[color=darkcyan,pattern=crosshatch,pattern color=darkcyan,node near coord style={font=\tiny,/pgf/number format/fixed, /pgf/number format/zerofill, /pgf/number format/precision=2, rotate=90, anchor=west}] coordinates {
                (1, 3.23) 
                (2, 2.21) 
                (3, 1.34) 
                (4, 1.13) 
            };
            \addlegendentry{Pot}
            \addplot[color=darkcyan,fill=darkcyan!30,node near coord style={font=\tiny,/pgf/number format/fixed, /pgf/number format/zerofill, /pgf/number format/precision=2}] coordinates {
                (1, 17.77) 
                (2, 11.74) 
                (3, 7.41 ) 
                (4, 6.21 ) 
            };
            \legend{}
        \end{axis}
    \end{tikzpicture}
    \caption{
	\textbf{How much faster is the execution of STMBench7 with $x$~threads if we want determinism?}
    The y axis measures the throughput using DeSTM, PoGL~(preordered global lock), Pot$-$~(ordered commits), Pot$*$~(ordered commits and transaction modes), and Pot~(ordered commits, transaction modes, and live promotion), normalized to the nondeterministic execution using the baseline STM~(higher is better).
    The titles indicate the workload type.
     }
    \label{fig:evaluation_perf_stm_sb7}
\end{figure*}
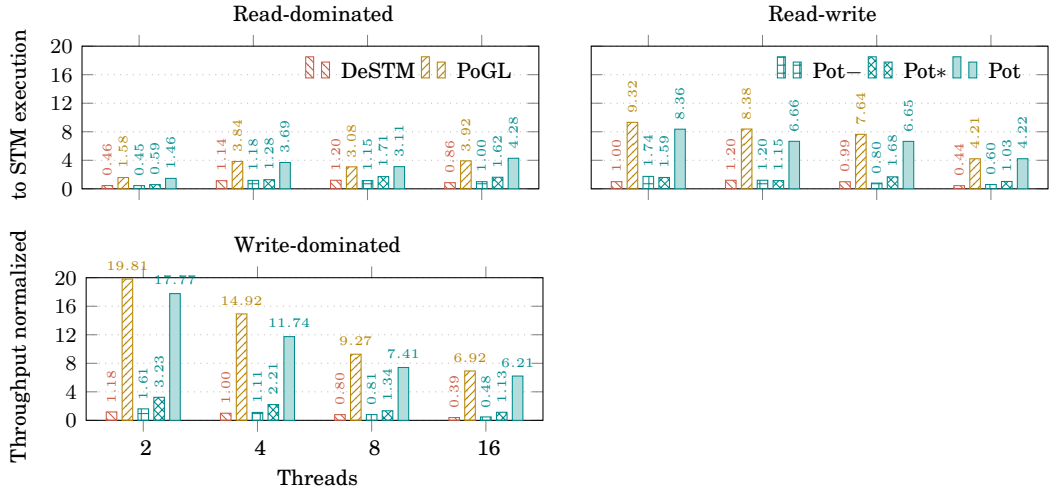
Fig.~\ref{fig:evaluation_perf_stm_stamp} quantifies the cost of deterministic multithreading when using DeSTM, PoGL, and Pot, on STAMP.
With it we seek to answer the following question: ``How much slower is the execution with $x$ threads if we want determinism?''
The Figure reports the execution time normalized to the baseline nondeterministic STM execution~(y~axis) of every benchmark of the STAMP suite, when executed with DeSTM, PoGL, Pot$-$~(ordered commits), Pot$*$~(ordered commits and transaction modes) and Pot~(ordered commits, transaction modes, and live promotion) using from 2 to 16 threads~(x~axis).
In these plots lower is better, and values below 1~mean that the deterministic execution was \emph{faster} than standard nondeterministic execution.
Four observations stand out: (a)~the cost of ensuring determinism increases with the number of threads, (b)~Pot outperforms DeSTM in all benchmarks, (c)~Pot is at most~${\approx 3 \times}$~slower than the nondeterministic baseline, while DeSTM suffers from a slowdown of up to~$\approx 11 \times$, (d)~Pot is even \emph{always faster} than the baseline STM execution on Genome, and (e)~although PoGL works well in some workloads, Pot achieves the best of both worlds: Pot is comparable to PoGL on the workloads PoGL works well, and considerably outperforms PoGL on the remaining workloads~(e.g. $\approx 2.5 \times$ on Intruder, $\approx 3 \times$ on Labyrinth and Vacation$+$, and $\approx 5 \times$ on Vacation$-$). 
\begin{figure}[tb]
    \centering
    \footnotesize
\begin{tabular}{rcccc}
\toprule
Benchmark & \multicolumn{4}{c}{Threads} \\
\cmidrule{2-5}     & 2             & 4             & 8            & 16            \\
\midrule
    Bayes          & $ 1.88\times$ & $ 1.03\times$ & $0.95\times$ & $ 0.68\times$ \\
    Genome         & $ 3.24\times$ & $ 3.99\times$ & $3.92\times$ & $ 3.24\times$ \\
    Intruder       & $ 2.92\times$ & $ 3.11\times$ & $2.19\times$ & $ 1.74\times$ \\
    Kmeans$-$      & $ 4.16\times$ & $ 2.54\times$ & $2.22\times$ & $ 1.68\times$ \\
    KMeans$+$      & $ 3.62\times$ & $ 2.76\times$ & $1.97\times$ & $ 1.16\times$ \\
    Labyrinth      & $ 6.61\times$ & $ 5.31\times$ & $2.67\times$ & $ 0.77\times$ \\
    SSCA2          & $ 4.29\times$ & $ 1.62\times$ & $1.36\times$ & $ 1.34\times$ \\
    Vacation$-$    & $ 5.52\times$ & $ 4.01\times$ & $3.51\times$ & $ 3.60\times$ \\
    Vacation$+$    & $ 5.91\times$ & $ 4.93\times$ & $4.41\times$ & $ 5.29\times$ \\
    Yada           & $ 3.00\times$ & $ 3.26\times$ & $2.23\times$ & $ 1.69\times$ \\
    STMBench7 (r)  & $ 2.90\times$ & $ 3.53\times$ & $2.34\times$ & $ 4.73\times$ \\
    STMBench7 (rw) & $ 8.02\times$ & $ 5.96\times$ & $7.16\times$ & $ 7.82\times$ \\
    STMBench7 (w)  & $15.10\times$ & $11.77\times$ & $9.17\times$ & $11.31\times$ \\
\bottomrule
\end{tabular}
    \caption{
    Time that DeSTM transactions spend waiting to enforce determinism compared to Pot, in STAMP.
    A value of~$2\times$ means that, on average, DeSTM transactions spend~$2\times$ more time waiting for their turn~(hence higher is better for Pot).
     }
    \label{fig:evaluation_stm_waiting2}
\end{figure}
\begin{figure}[tb]
	\centering
	\begin{subfigure}[t]{0.3\linewidth}
		\centering
		\includegraphics[width=0.6\linewidth]{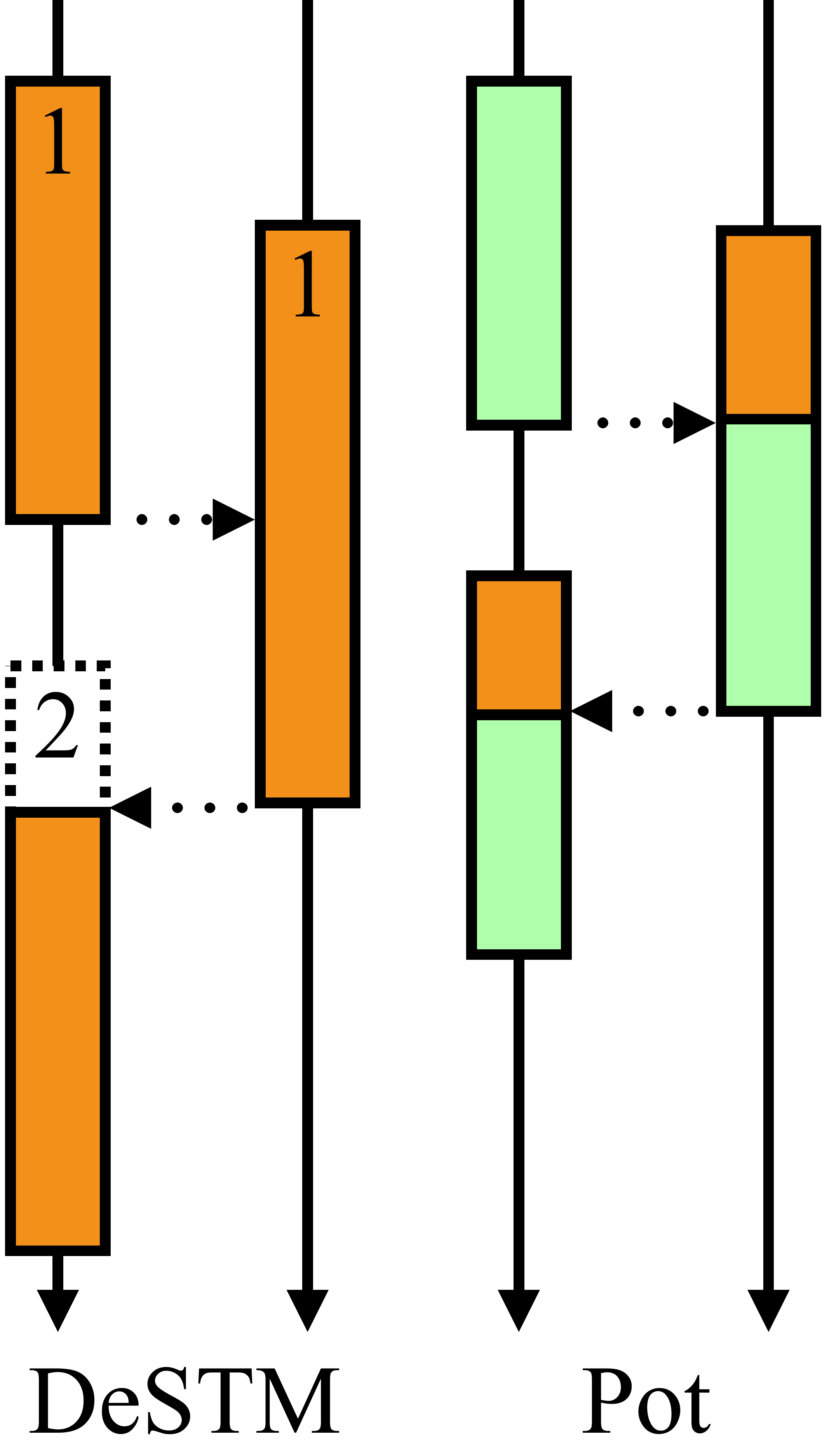}
		\caption{DeSTM waits to start.}
		\label{subfig:evaluation_stm_pot_vs_destm_start}
	\end{subfigure}
	\begin{subfigure}[t]{0.3\linewidth}
		\centering
		\includegraphics[width=0.6\linewidth]{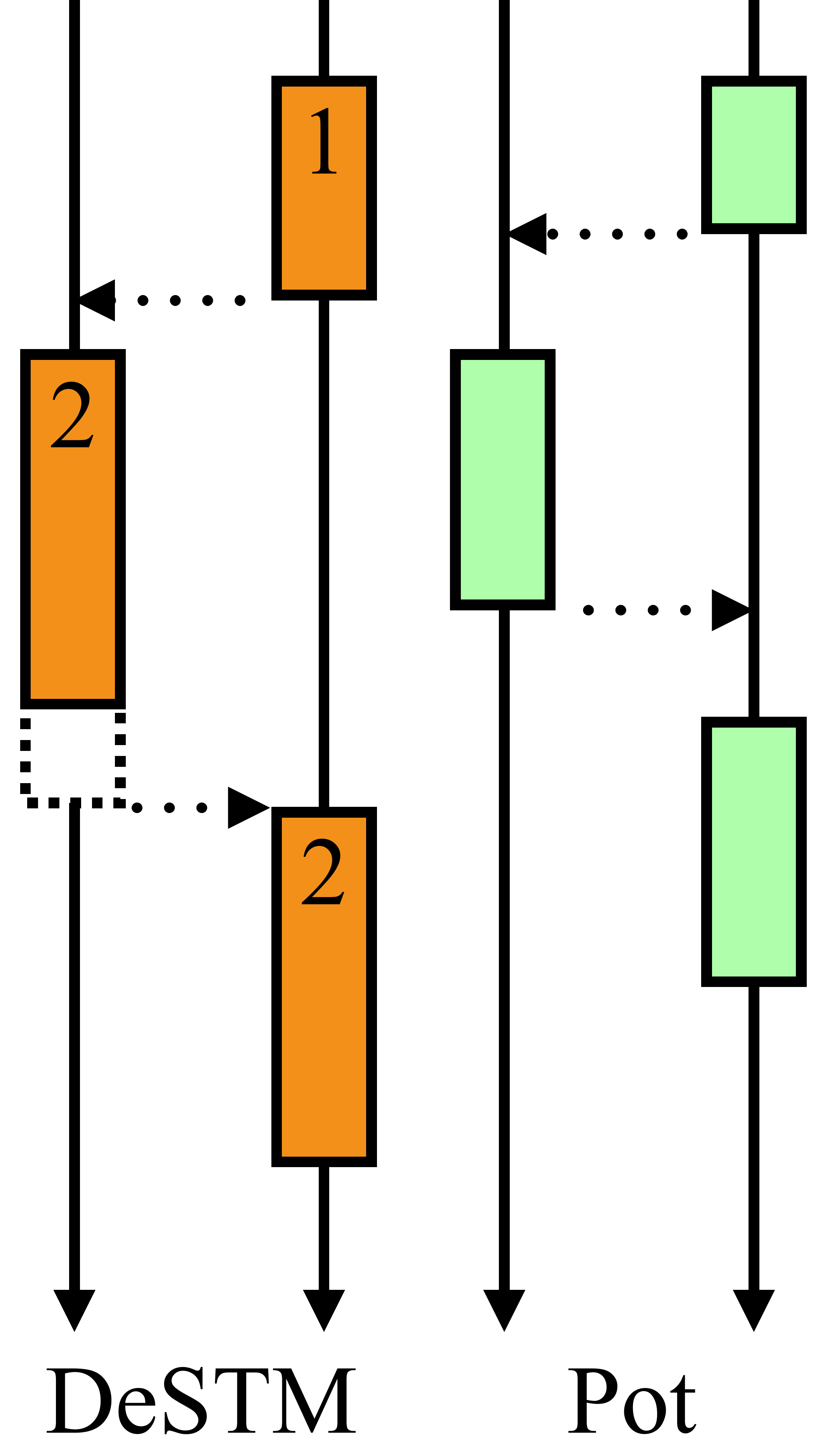}
		\caption{DeSTM waits to commit.}
		\label{subfig:evaluation_stm_pot_vs_destm_commit}
	\end{subfigure}
	\begin{subfigure}[t]{0.3\linewidth}
		\includegraphics[width=0.8\linewidth]{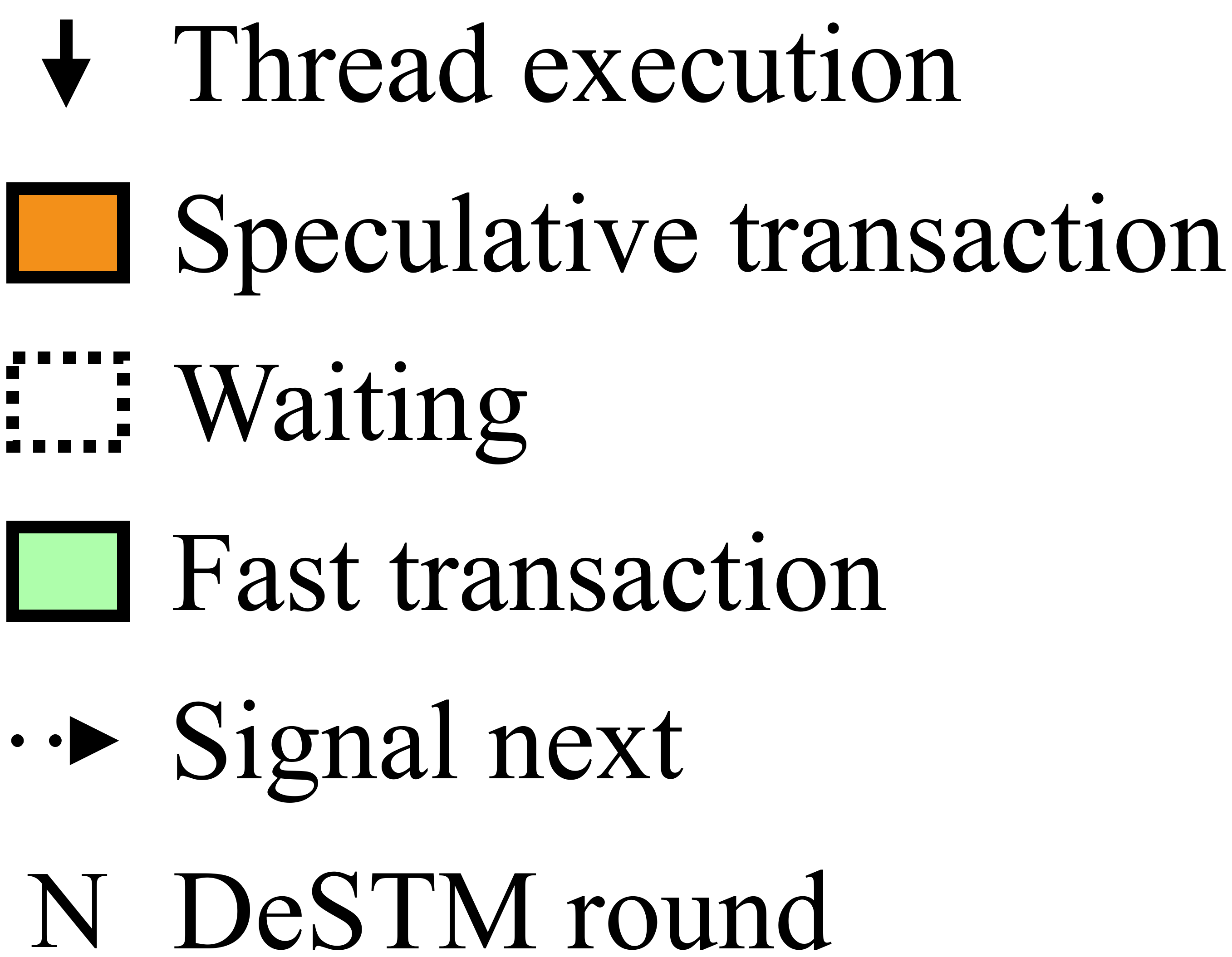}
	\end{subfigure}
    \caption{
        Examples of the difference between DeSTM and Pot.
        In DeSTM time is divided into rounds, and in each round each thread executes one transaction. A transaction cannot start if some transaction from the previous round has not finished yet (a), and cannot commit, even on its turn, if some transaction from the same round has not started yet (b).
        In contrast, Pot realizes that rounds are not necessary to respect a predefined serial order, so transactions never wait to start, nor to commit on their turn.
        Pot also accelerates the execution of the next transaction to commit according to the serial order.
    }
    \label{fig:evaluation_stm_pot_vs_destm}
\end{figure}

The fact that the cost of ensuring determinism increases with the number of threads is unsurprising; the probability of a transaction~$t$~attempting to commit before its turn increases with the number of threads, particularly if there are transactions ordered before~$t$ that take longer than~$t$.
Pot's ordered commits and transaction modes minimize these situations to increase the probability of transactions not having to wait for their turn to commit.
Fig.~\ref{fig:evaluation_stm_waiting2} supports this claim.
It shows, for each benchmark/thread combination, how much time DeSTM transactions ``waste'' to enforce determinism, on average, when compared to Pot.
We can observe that in general DeSTM transactions spend more time waiting for their turn to commit.
Fig.~\ref{fig:evaluation_stm_pot_vs_destm} shows two example scenarios that highlight the differences between DeSTM and Pot.
In DeSTM time is divided into rounds, and in each round each thread executes one transaction. A transaction cannot start if some transaction from the previous round has not finished yet (Fig.~\ref{subfig:evaluation_stm_pot_vs_destm_start}), and cannot commit, even on its turn, if some transaction from the same round has not started yet (Fig.~\ref{subfig:evaluation_stm_pot_vs_destm_commit}).
In contrast, Pot realizes that rounds are not necessary to respect a predefined serial order, so transactions never wait to start, nor to commit on their turn.

Pot also accelerates the execution of the next transaction to commit according to the serial order.
From Fig.~\ref{fig:evaluation_stm_fast} we deduce that the benefits of the fast mode should be more apparent in benchmarks with bigger transactions with higher write-to-read ratio, and/or higher contention.
Fast transactions~(Pot$*$) further improve performance over ordered commits in all benchmarks~(Fig.~\ref{fig:evaluation_perf_stm_stamp}).
However, the overhead of our implementation of live promotion~(Pot) only pays off in Genome, Vacation$+$, and Yada.

We also experimented with STMBench7.
Fig.~\ref{fig:evaluation_perf_stm_sb7} shows the throughput of DeSTM, PoGL, Pot$-$, Pot$*$, and Pot, normalized to the throughput of the baseline STM.
Because STMBench7 features a more diverse set of transaction profiles, with more complex read-write transactions, live promotion is very effective at boosting Pot's throughput: in fact, Pot is \emph{always faster} than the nondeterministic baseline, usually by more than~$3 \times$.

To conclude, in our experiments Pot is a good general solution because it achieves the best of both worlds: when speculation is effective Pot provides superior performance, and when speculation is not effective Pot's performance is very close to PoGL's~(Figs.~\ref{fig:evaluation_perf_stm_stamp}~and~\ref{fig:evaluation_perf_stm_sb7}).
Pot's excellent results compared to DeSTM's are explained by both the speedups that fast transaction can achieve, as observed in Fig.~\ref{fig:evaluation_stm_fast}, and the decrease of the time transactions spend waiting for their turn, as we observe in Fig.~\ref{fig:evaluation_stm_waiting2}.
Pot marks a significant advance over the state of the art in performance, and provides promising evidence that using both \acrshort{stm} and determinism to enable multithreaded replicas for fault tolerance, and/or to ease multithreaded programming, may be practical.

\noindent{\textbf{Scalability.}} 
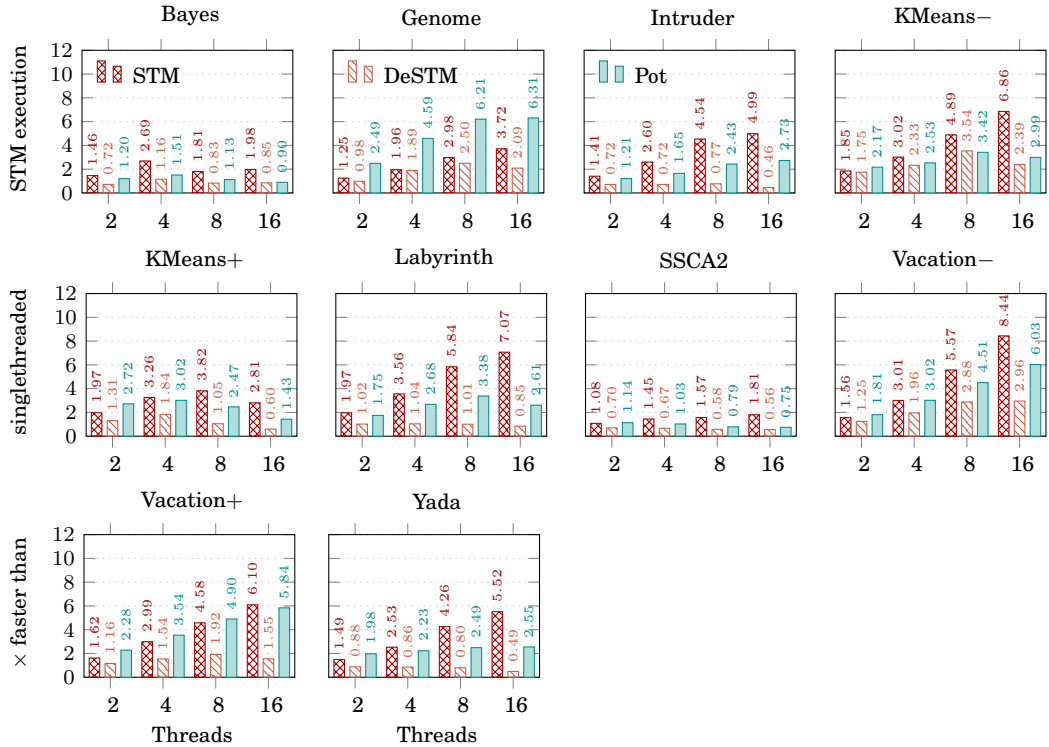
\begin{figure*}[tb]
    \begin{tikzpicture}
        \begin{axis}[
            xmin=0.5, xmax=4.5,
            xtick={1,2,3,4},
            xticklabels={2,4,8,16},
            x tick label style={font=\footnotesize},
            x label style={font=\footnotesize},
            ymin=0, ymax=12,
            ytick={0,2,4,6,8,10,12},
            ymajorgrids=true,
            y tick label style={font=\footnotesize},
            y label style={align=left,font=\footnotesize},
            ylabel near ticks,
            ylabel={STM execution},
            ylabel style={align=center,font=\footnotesize},
            legend style={font=\footnotesize,fill=none,draw=none},
            legend pos=north west,
            title style={font=\footnotesize},
            grid style={dotted,gray!50},
            width=0.315\textwidth,
            height=0.25\textwidth,
            ybar,
            bar width=4pt,
            nodes near coords,
            every node near coord/.append style={font=\tiny,/pgf/number format/fixed, /pgf/number format/zerofill, /pgf/number format/precision=2, rotate=90, anchor=west},
            title={Bayes},
        ]
            \addlegendentry{TL2}
            \addplot[color=darkcandyapplered,pattern=crosshatch,pattern color=darkcandyapplered] coordinates {
                (1, 1.46) 
                (2, 2.69) 
                (3, 1.81) 
                (4, 1.98) 
            };
            \addlegendentry{DeSTM}
            \addplot[color=darkcoral,pattern=north west lines,pattern color=darkcoral] coordinates {
                (1, .72 ) 
                (2, 1.16) 
                (3, .83 ) 
                (4, .85 ) 
            };
            \addlegendentry{Pot}
            \addplot[color=darkcyan,fill=darkcyan!30] coordinates {
                (1, 1.20) 
                (2, 1.51) 
                (3, 1.13) 
                (4, .90 ) 
            };
            \legend{STM,,}
        \end{axis}
    \end{tikzpicture}
    \begin{tikzpicture}
        \begin{axis}[
            xmin=0.5, xmax=4.5,
            xtick={1,2,3,4},
            xticklabels={2,4,8,16},
            x tick label style={font=\footnotesize},
            x label style={font=\footnotesize},
            ymin=0, ymax=12,
            ytick={0,2,4,6,8,10,12},
            yticklabels={,,,,,},
            ymajorgrids=true,
            y tick label style={font=\footnotesize},
            y label style={align=left,font=\footnotesize},
            ylabel near ticks,
            ylabel style={align=center,font=\footnotesize},
            legend style={font=\footnotesize,fill=none,draw=none},
            legend pos=north west,
            title style={font=\footnotesize},
            grid style={dotted,gray!50},
            width=0.315\textwidth,
            height=0.25\textwidth,
            ybar,
            bar width=4pt,
            nodes near coords,
            every node near coord/.append style={font=\tiny,/pgf/number format/fixed, /pgf/number format/zerofill, /pgf/number format/precision=2, rotate=90, anchor=west},
            title={Genome},
        ]
            \addlegendentry{TL2}
            \addplot[color=darkcandyapplered,pattern=crosshatch,pattern color=darkcandyapplered] coordinates {
                (1, 1.25) 
                (2, 1.96) 
                (3, 2.98) 
                (4, 3.72) 
            };
            \addlegendentry{DeSTM}
            \addplot[color=darkcoral,pattern=north west lines,pattern color=darkcoral] coordinates {
                (1, .98 ) 
                (2, 1.89) 
                (3, 2.50) 
                (4, 2.09) 
            };
            \addlegendentry{Pot}
            \addplot[color=darkcyan,fill=darkcyan!30] coordinates {
                (1, 2.49) 
                (2, 4.59) 
                (3, 6.21) 
                (4, 6.31) 
            };
            \legend{,DeSTM,}
        \end{axis}
    \end{tikzpicture}
    \begin{tikzpicture}
        \begin{axis}[
            xmin=0.5, xmax=4.5,
            xtick={1,2,3,4},
            xticklabels={2,4,8,16},
            x tick label style={font=\footnotesize},
            x label style={font=\footnotesize},
            ymin=0, ymax=12,
            ytick={0,2,4,6,8,10,12},
            yticklabels={,,,,,},
            ymajorgrids=true,
            y tick label style={font=\footnotesize},
            y label style={align=left,font=\footnotesize},
            ylabel near ticks,
            ylabel style={align=center,font=\footnotesize},
            legend style={font=\footnotesize,fill=none,draw=none},
            legend pos=north west,
            title style={font=\footnotesize},
            grid style={dotted,gray!50},
            width=0.315\textwidth,
            height=0.25\textwidth,
            ybar,
            bar width=4pt,
            nodes near coords,
            every node near coord/.append style={font=\tiny,/pgf/number format/fixed, /pgf/number format/zerofill, /pgf/number format/precision=2, rotate=90, anchor=west},
            title={Intruder},
        ]
            \addlegendentry{TL2}
            \addplot[color=darkcandyapplered,pattern=crosshatch,pattern color=darkcandyapplered] coordinates {
                (1, 1.41) 
                (2, 2.60) 
                (3, 4.54) 
                (4, 4.99) 
            };
            \addlegendentry{DeSTM}
            \addplot[color=darkcoral,pattern=north west lines,pattern color=darkcoral] coordinates {
                (1, .72) 
                (2, .72) 
                (3, .77) 
                (4, .46) 
            };
            \addlegendentry{Pot}
            \addplot[color=darkcyan,fill=darkcyan!30] coordinates {
                (1, 1.21) 
                (2, 1.65) 
                (3, 2.43) 
                (4, 2.73) 
            };
            \legend{,,Pot}
        \end{axis}
    \end{tikzpicture}
    \begin{tikzpicture}
        \begin{axis}[
            xmin=0.5, xmax=4.5,
            xtick={1,2,3,4},
            xticklabels={2,4,8,16},
            x tick label style={font=\footnotesize},
            x label style={font=\footnotesize},
            ymin=0, ymax=12,
            ytick={0,2,4,6,8,10,12},
            yticklabels={,,,,,},
            ymajorgrids=true,
            y tick label style={font=\footnotesize},
            y label style={align=left,font=\footnotesize},
            ylabel near ticks,
            ylabel style={align=center,font=\footnotesize},
            legend style={font=\footnotesize,fill=none,draw=none},
            legend pos=north west,
            title style={font=\footnotesize},
            grid style={dotted,gray!50},
            width=0.315\textwidth,
            height=0.25\textwidth,
            ybar,
            bar width=4pt,
            nodes near coords,
            every node near coord/.append style={font=\tiny,/pgf/number format/fixed, /pgf/number format/zerofill, /pgf/number format/precision=2, rotate=90, anchor=west},
            title={KMeans$-$},
        ]
            \addlegendentry{TL2}
            \addplot[color=darkcandyapplered,pattern=crosshatch,pattern color=darkcandyapplered] coordinates {
                (1, 1.85) 
                (2, 3.02) 
                (3, 4.89) 
                (4, 6.86) 
            };
            \addlegendentry{DeSTM}
            \addplot[color=darkcoral,pattern=north west lines,pattern color=darkcoral] coordinates {
                (1, 1.75) 
                (2, 2.33) 
                (3, 3.54) 
                (4, 2.39) 
            };
            \addlegendentry{Pot}
            \addplot[color=darkcyan,fill=darkcyan!30] coordinates {
                (1, 2.17) 
                (2, 2.53) 
                (3, 3.42) 
                (4, 2.99) 
            };
            \legend{}
        \end{axis}
    \end{tikzpicture}
    \begin{tikzpicture}
        \begin{axis}[
            xmin=0.5, xmax=4.5,
            xtick={1,2,3,4},
            xticklabels={2,4,8,16},
            x tick label style={font=\footnotesize},
            x label style={font=\footnotesize},
            ymin=0, ymax=12,
            ytick={0,2,4,6,8,10,12},
            ymajorgrids=true,
            y tick label style={font=\footnotesize},
            y label style={align=left,font=\footnotesize},
            ylabel near ticks,
            ylabel={singlethreaded},
            ylabel style={align=center,font=\footnotesize},
            legend style={font=\footnotesize,fill=none,draw=none},
            legend pos=north west,
            title style={font=\footnotesize},
            grid style={dotted,gray!50},
            width=0.315\textwidth,
            height=0.25\textwidth,
            ybar,
            bar width=4pt,
            nodes near coords,
            every node near coord/.append style={font=\tiny,/pgf/number format/fixed, /pgf/number format/zerofill, /pgf/number format/precision=2, rotate=90, anchor=west},
            title={KMeans$+$},
        ]
            \addlegendentry{TL2}
            \addplot[color=darkcandyapplered,pattern=crosshatch,pattern color=darkcandyapplered] coordinates {
                (1, 1.97) 
                (2, 3.26) 
                (3, 3.82) 
                (4, 2.81) 
            };
            \addlegendentry{DeSTM}
            \addplot[color=darkcoral,pattern=north west lines,pattern color=darkcoral] coordinates {
                (1, 1.31) 
                (2, 1.84) 
                (3, 1.05) 
                (4, .60 ) 
            };
            \addlegendentry{Pot}
            \addplot[color=darkcyan,fill=darkcyan!30] coordinates {
                (1, 2.72) 
                (2, 3.02) 
                (3, 2.47) 
                (4, 1.43) 
            };
            \legend{}
        \end{axis}
    \end{tikzpicture}
    \begin{tikzpicture}
        \begin{axis}[
            xmin=0.5, xmax=4.5,
            xtick={1,2,3,4},
            xticklabels={2,4,8,16},
            x tick label style={font=\footnotesize},
            x label style={font=\footnotesize},
            ymin=0, ymax=12,
            ytick={0,2,4,6,8,10,12},
            yticklabels={,,,,,},
            ymajorgrids=true,
            y tick label style={font=\footnotesize},
            y label style={align=left,font=\footnotesize},
            ylabel near ticks,
            ylabel style={align=center,font=\footnotesize},
            legend style={font=\footnotesize,fill=none,draw=none},
            legend pos=north west,
            title style={font=\footnotesize},
            grid style={dotted,gray!50},
            width=0.315\textwidth,
            height=0.25\textwidth,
            ybar,
            bar width=4pt,
            nodes near coords,
            every node near coord/.append style={font=\tiny,/pgf/number format/fixed, /pgf/number format/zerofill, /pgf/number format/precision=2, rotate=90, anchor=west},
            title={Labyrinth},
        ]
            \addlegendentry{TL2}
            \addplot[color=darkcandyapplered,pattern=crosshatch,pattern color=darkcandyapplered] coordinates {
                (1, 1.97) 
                (2, 3.56) 
                (3, 5.84) 
                (4, 7.07) 
            };
            \addlegendentry{DeSTM}
            \addplot[color=darkcoral,pattern=north west lines,pattern color=darkcoral] coordinates {
                (1, 1.02) 
                (2, 1.04) 
                (3, 1.01) 
                (4, .85 ) 
            };
            \addlegendentry{Pot}
            \addplot[color=darkcyan,fill=darkcyan!30] coordinates {
                (1, 1.75) 
                (2, 2.68) 
                (3, 3.38) 
                (4, 2.61) 
            };
            \legend{}
        \end{axis}
    \end{tikzpicture}
    \begin{tikzpicture}
        \begin{axis}[
            xmin=0.5, xmax=4.5,
            xtick={1,2,3,4},
            xticklabels={2,4,8,16},
            x tick label style={font=\footnotesize},
            x label style={font=\footnotesize},
            ymin=0, ymax=12,
            ytick={0,2,4,6,8,10,12},
            yticklabels={,,,,,},
            ymajorgrids=true,
            y tick label style={font=\footnotesize},
            y label style={align=left,font=\footnotesize},
            ylabel near ticks,
            ylabel style={align=center,font=\footnotesize},
            legend style={font=\footnotesize,fill=none,draw=none},
            legend pos=north west,
            title style={font=\footnotesize},
            grid style={dotted,gray!50},
            width=0.315\textwidth,
            height=0.25\textwidth,
            ybar,
            bar width=4pt,
            nodes near coords,
            every node near coord/.append style={font=\tiny,/pgf/number format/fixed, /pgf/number format/zerofill, /pgf/number format/precision=2, rotate=90, anchor=west},
            title={SSCA2},
        ]
            \addlegendentry{TL2}
            \addplot[color=darkcandyapplered,pattern=crosshatch,pattern color=darkcandyapplered] coordinates {
                (1, 1.08) 
                (2, 1.45) 
                (3, 1.57) 
                (4, 1.81) 
            };
            \addlegendentry{DeSTM}
            \addplot[color=darkcoral,pattern=north west lines,pattern color=darkcoral] coordinates {
                (1, .70) 
                (2, .67) 
                (3, .58) 
                (4, .56) 
            };
            \addlegendentry{Pot}
            \addplot[color=darkcyan,fill=darkcyan!30] coordinates {
                (1, 1.14) 
                (2, 1.03) 
                (3, .79 ) 
                (4, .75 ) 
            };
            \legend{}
        \end{axis}
    \end{tikzpicture}
    \begin{tikzpicture}
        \begin{axis}[
            xmin=0.5, xmax=4.5,
            xtick={1,2,3,4},
            xticklabels={2,4,8,16},
            x tick label style={font=\footnotesize},
            x label style={font=\footnotesize},
            ymin=0, ymax=12,
            ytick={0,2,4,6,8,10,12},
            yticklabels={,,,,,},
            ymajorgrids=true,
            y tick label style={font=\footnotesize},
            y label style={align=left,font=\footnotesize},
            ylabel near ticks,
            ylabel style={align=center,font=\footnotesize},
            legend style={font=\footnotesize,fill=none,draw=none},
            legend pos=north west,
            title style={font=\footnotesize},
            grid style={dotted,gray!50},
            width=0.315\textwidth,
            height=0.25\textwidth,
            ybar,
            bar width=4pt,
            nodes near coords,
            every node near coord/.append style={font=\tiny,/pgf/number format/fixed, /pgf/number format/zerofill, /pgf/number format/precision=2, rotate=90, anchor=west},
            title={Vacation$-$},
        ]
            \addlegendentry{TL2}
            \addplot[color=darkcandyapplered,pattern=crosshatch,pattern color=darkcandyapplered] coordinates {
                (1, 1.56) 
                (2, 3.01) 
                (3, 5.57) 
                (4, 8.44) 
            };
            \addlegendentry{DeSTM}
            \addplot[color=darkcoral,pattern=north west lines,pattern color=darkcoral] coordinates {
                (1, 1.25) 
                (2, 1.96) 
                (3, 2.88) 
                (4, 2.96) 
            };
            \addlegendentry{Pot}
            \addplot[color=darkcyan,fill=darkcyan!30] coordinates {
                (1, 1.81) 
                (2, 3.02) 
                (3, 4.51) 
                (4, 6.03) 
            };
            \legend{}
        \end{axis}
    \end{tikzpicture}
    \begin{tikzpicture}
        \begin{axis}[
            xmin=0.5, xmax=4.5,
            xtick={1,2,3,4},
            xticklabels={2,4,8,16},
            x tick label style={font=\footnotesize},
            x label style={font=\footnotesize},
            xlabel={Threads},
            ymin=0, ymax=12,
            ytick={0,2,4,6,8,10,12},
            ymajorgrids=true,
            y tick label style={font=\footnotesize},
            y label style={align=left,font=\footnotesize},
            ylabel near ticks,
            ylabel={$\times$ faster than},
            ylabel style={align=center,font=\footnotesize},
            legend style={font=\footnotesize,fill=none,draw=none},
            legend pos=north west,
            title style={font=\footnotesize},
            grid style={dotted,gray!50},
            width=0.315\textwidth,
            height=0.25\textwidth,
            ybar,
            bar width=4pt,
            nodes near coords,
            every node near coord/.append style={font=\tiny,/pgf/number format/fixed, /pgf/number format/zerofill, /pgf/number format/precision=2, rotate=90, anchor=west},
            title={Vacation$+$},
        ]
            \addlegendentry{TL2}
            \addplot[color=darkcandyapplered,pattern=crosshatch,pattern color=darkcandyapplered] coordinates {
                (1, 1.62) 
                (2, 2.99) 
                (3, 4.58) 
                (4, 6.10) 
            };
            \addlegendentry{DeSTM}
            \addplot[color=darkcoral,pattern=north west lines,pattern color=darkcoral] coordinates {
                (1, 1.16) 
                (2, 1.54) 
                (3, 1.92) 
                (4, 1.55) 
            };
            \addlegendentry{Pot}
            \addplot[color=darkcyan,fill=darkcyan!30] coordinates {
                (1, 2.28) 
                (2, 3.54) 
                (3, 4.90) 
                (4, 5.84) 
            };
            \legend{}
        \end{axis}
    \end{tikzpicture}
    \begin{tikzpicture}
        \begin{axis}[
            xmin=0.5, xmax=4.5,
            xtick={1,2,3,4},
            xticklabels={2,4,8,16},
            x tick label style={font=\footnotesize},
            x label style={font=\footnotesize},
            xlabel={Threads},
            ymin=0, ymax=12,
            ytick={0,2,4,6,8,10,12},
            yticklabels={,,,,,},
            ymajorgrids=true,
            y tick label style={font=\footnotesize},
            y label style={align=left,font=\footnotesize},
            ylabel near ticks,
            ylabel style={align=center,font=\footnotesize},
            legend style={font=\footnotesize,fill=none,draw=none},
            legend pos=north west,
            title style={font=\footnotesize},
            grid style={dotted,gray!50},
            width=0.315\textwidth,
            height=0.25\textwidth,
            ybar,
            bar width=4pt,
            nodes near coords,
            every node near coord/.append style={font=\tiny,/pgf/number format/fixed, /pgf/number format/zerofill, /pgf/number format/precision=2, rotate=90, anchor=west},
            title={Yada},
        ]
            \addlegendentry{TL2}
            \addplot[color=darkcandyapplered,pattern=crosshatch,pattern color=darkcandyapplered] coordinates {
                (1, 1.49) 
                (2, 2.53) 
                (3, 4.26) 
                (4, 5.52) 
            };
            \addlegendentry{DeSTM}
            \addplot[color=darkcoral,pattern=north west lines,pattern color=darkcoral] coordinates {
                (1, .88) 
                (2, .86) 
                (3, .80) 
                (4, .49) 
            };
            \addlegendentry{Pot}
            \addplot[color=darkcyan,fill=darkcyan!30] coordinates {
                (1, 1.98) 
                (2, 2.23) 
                (3, 2.49) 
                (4, 2.55) 
            };
            \legend{}
        \end{axis}
    \end{tikzpicture}
    \caption{
        Scalability of deterministic execution using DeSTM and Pot on STAMP.
        The y axis measures the speedup over a singlethread baseline STM execution. 
        A value of 1 means the execution time was the same as the baseline, a value greater than 1 means the execution time was faster (better), and a value less than 1 means the execution time was slower (worse).
    }
    \label{fig:evaluation_scalability_stm_stamp}
\end{figure*}
\begin{figure*}[tb]
    \centering
    \begin{tikzpicture}
        \begin{axis}[
            xmin=0.5, xmax=4.5,
            xtick={1,2,3,4},
            xticklabels={2,4,8,16},
            x tick label style={font=\footnotesize},
            x label style={font=\footnotesize},
            xlabel={Threads},
            ymin=0, ymax=20,
            ytick={0,4,8,12,16,20},
            ymajorgrids=true,
            y tick label style={font=\footnotesize},
            y label style={align=left,font=\footnotesize},
            ylabel near ticks,
            ylabel={$\times$ faster than \\ singlethreaded \\ STM execution},
            ylabel style={align=center,font=\footnotesize},
            legend columns=1,
            legend style={font=\footnotesize,fill=none,draw=none},
            legend pos=north west,
            title={Read-dominated},
            title style={font=\footnotesize},
            grid style={dotted,gray!50},
            width=0.38\textwidth,
            height=0.25\textwidth,
            ybar,
            bar width=4pt,
            nodes near coords,
            every node near coord/.append style={font=\tiny,/pgf/number format/fixed, /pgf/number format/zerofill, /pgf/number format/precision=2, rotate=90, anchor=west}
        ]
            \addlegendentry{TL2}
            \addplot[color=darkcandyapplered,pattern=crosshatch,pattern color=darkcandyapplered] coordinates {
                (1, 1.43) 
                (2, .73 ) 
                (3, .91 ) 
                (4, .65 ) 
            };
            \addlegendentry{DeSTM}
            \addplot[color=darkcoral,pattern=north west lines,pattern color=darkcoral] coordinates {
                (1, .66 ) 
                (2, .84 ) 
                (3, 1.10) 
                (4, .57 ) 
            };
            \addlegendentry{Pot}
            \addplot[color=darkcyan,fill=darkcyan!30] coordinates {
                (1, 2.11) 
                (2, 2.71) 
                (3, 2.84) 
                (4, 2.82) 
            };
            \legend{STM,,}
        \end{axis}
    \end{tikzpicture}
    \begin{tikzpicture}
        \begin{axis}[
            xmin=0.5, xmax=4.5,
            xtick={1,2,3,4},
            xticklabels={2,4,8,16},
            x tick label style={font=\footnotesize},
            x label style={font=\footnotesize},
            xlabel={Threads},
            ymin=0, ymax=20,
            ytick={0,4,8,12,16,20},
            yticklabels={,,,,,,},
            ymajorgrids=true,
            y tick label style={font=\footnotesize},
            y label style={align=center,font=\footnotesize},
            ylabel near ticks,
            ylabel style={align=center,font=\footnotesize},
            legend columns=2,
            legend style={font=\footnotesize,fill=none,draw=none},
            legend pos=north west,
            title={Read-write},
            title style={font=\footnotesize},
            grid style={dotted,gray!50},
            width=0.38\textwidth,
            height=0.25\textwidth,
            ybar,
            bar width=4pt,
            nodes near coords,
            every node near coord/.append style={font=\tiny,/pgf/number format/fixed, /pgf/number format/zerofill, /pgf/number format/precision=2, rotate=90, anchor=west}
        ]
            \addlegendentry{TL2}
            \addplot[color=darkcandyapplered,pattern=crosshatch,pattern color=darkcandyapplered] coordinates {
                (1, .93 ) 
                (2, .96 ) 
                (3, 1.05) 
                (4, 1.47) 
            };
            \addlegendentry{DeSTM}
            \addplot[color=darkcoral,pattern=north west lines,pattern color=darkcoral] coordinates {
                (1, .93 ) 
                (2, 1.16) 
                (3, 1.04) 
                (4, .66 ) 
            };
            \addlegendentry{Pot}
            \addplot[color=darkcyan,fill=darkcyan!30] coordinates {
                (1, 7.78) 
                (2, 6.45) 
                (3, 7.03) 
                (4, 6.22) 
            };
            \legend{,DeSTM,Pot}
        \end{axis}
    \end{tikzpicture}
    \begin{tikzpicture}
        \begin{axis}[
            xmin=0.5, xmax=4.5,
            xtick={1,2,3,4},
            xticklabels={2,4,8,16},
            x tick label style={font=\footnotesize},
            x label style={font=\footnotesize},
            xlabel={Threads},
            ymin=0, ymax=20,
            ytick={0,4,8,12,16,20},
            yticklabels={,,,,,,},
            ymajorgrids=true,
            y tick label style={font=\footnotesize},
            y label style={align=center,font=\footnotesize},
            ylabel near ticks,
            ylabel style={align=center,font=\footnotesize},
            legend style={font=\footnotesize,fill=none,draw=none},
            legend pos=north west,
            title={Write-dominated},
            title style={font=\footnotesize},
            grid style={dotted,gray!50},
            width=0.38\textwidth,
            height=0.25\textwidth,
            ybar,
            bar width=4pt,
            nodes near coords,
        ]
            \addlegendentry{TL2}
            \addplot[color=darkcandyapplered,pattern=crosshatch,pattern color=darkcandyapplered,node near coord style={font=\tiny,/pgf/number format/fixed, /pgf/number format/zerofill, /pgf/number format/precision=2, rotate=90, anchor=west}] coordinates {
                (1, 1.10) 
                (2, 1.27) 
                (3, 1.52) 
                (4, 2.27) 
            };
            \addlegendentry{DeSTM}
            \addplot[color=darkcoral,pattern=north west lines,pattern color=darkcoral,node near coord style={font=\tiny,/pgf/number format/fixed, /pgf/number format/zerofill, /pgf/number format/precision=2, rotate=90, anchor=west}] coordinates {
                (1, 1.30) 
                (2, 1.27) 
                (3, 1.22) 
                (4, .89 ) 
            };
            \addlegendentry{Pot}
            \addplot[color=darkcyan,fill=darkcyan!30,node near coord style={font=\tiny,/pgf/number format/fixed, /pgf/number format/zerofill, /pgf/number format/precision=2}] coordinates {
                (1, 19.55) 
                (2, 14.92) 
                (3, 11.31) 
                (4, 14.12) 
            };
            \legend{}
        \end{axis}
    \end{tikzpicture}
    \caption{
    Scalability of deterministic execution using DeSTM and Pot on STMBench7.
    %
    %
    The y axis measures the speedup over a singlethreaded baseline STM execution.
    A value of 1 means the throughput was the same as the baseline, a value greater than 1 means the throughput was greater (better), and a value less than 1 means the throughput was lower (worse).
    The titles indicate the workload type.
     }
    \label{fig:evaluation_scalability_stm_sb7}
\end{figure*}
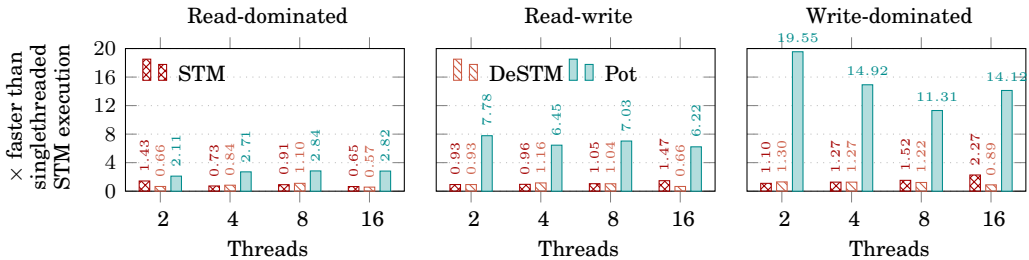
We further evaluate Pot's scalability compared to a singlethread execution using the baseline STM on STMBench7 and all of the STAMP benchmarks.
For comparison we also show results for DeSTM and the baseline STM itself.
The baseline's behavior serves as a guide for what to expect from Pot and DeSTM's implementation: we don't expect them to scale if the baseline does not scale.
However, ideally we should expect the Pot and DeSTM implementation to scale, even if shyly, despite the overheads required to ensure determinism, particularly the need to wait to enforce the deterministic commit order.
Figs.~\ref{fig:evaluation_scalability_stm_stamp} and~\ref{fig:evaluation_scalability_stm_sb7} show the results for STAMP and STMBench7, respectively.
We observe that DeSTM fails to scale, whereas Pot is able to scale up to some point, notably in Genome, Intruder and Vacation.
Pot shows better results than the baseline on STMBench7 because Pot inherently provides stronger progress guarantees to the more complex transactions in the benchmark: while they struggle to commit in the baseline STM, in Pot they eventually do when it is their turn, and even have their execution sped up by the fast mode.

As threads increase it becomes increasingly challenging to mask the overhead required to ensure determinism, but nonetheless Pot manages to keep up with the baseline up to a point.
As part of future work we plan to address this issue by taking advantage of commutativity: if two successive transactions in the predefined serial order commute they can both execute simultaneously as fast transactions.
The knowledge of whether two transactions commute can either be fed by the programmer via some sort of annotations, or inferred via analysis.
\subsection{\Acrlong{htm}} 
\label{sub:evaluation_htm}
We also evaluate our Pot \acrshort{htm} implementation using the STAMP benchmark suite.
We are interested in answering the following questions: (1)~how effective are fast transactions,~and (2)~what is the cost that Pot incurs in to ensure deterministic execution.

\noindent{\textbf{Are fast transactions effective?~(\S\ref{subsub:evaluation_htm_fast}.)}}
Yes, Pot fast transactions enjoy increased capacity limits when compared to regular transactions.
Our experiments show that for~4~of the STAMP benchmarks, Pot fast transactions greatly reduce the need to fall back to software~(Fig.~\ref{fig:evaluation_htm_hwsuccess}).

\noindent{\textbf{What is the cost that Pot incurs in to ensure determinism?~(\S\ref{subsub:evaluation_htm_perf}.)}}
Our experiments show that Pot ensures deterministic execution across all of STAMP's benchmarks with moderate overhead~(Fig.~\ref{fig:evaluation_htm_perf}.)
\subsubsection{Effectiveness of fast transactions} 
\label{subsub:evaluation_htm_fast}
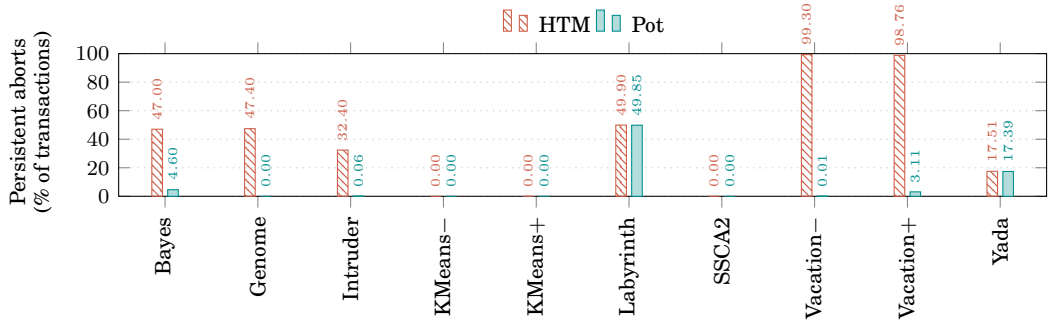
\begin{figure}[tb]
    \centering
    \begin{tikzpicture}
        \begin{axis}[
            xmin=0.5, xmax=10.5,
            xtick={1,2,3,4,5,6,7,8,9,10},
            xticklabels={Bayes,Genome,Intruder,KMeans$-$,KMeans$+$,Labyrinth,SSCA2,Vacation$-$,Vacation$+$,Yada},
            x label style={font=\footnotesize},
            x tick label style={font=\footnotesize,rotate=90,anchor=east},
            ymin=0, ymax=100,
            ytick={0,20,40,60,80,100},
            ymajorgrids=true,
            y tick label style={font=\footnotesize},            
            ylabel near ticks,
            ylabel={Persistent aborts \\ (\% of transactions)},
            ylabel style={align=center,font=\footnotesize},
            legend columns=2,
            legend style={font=\footnotesize,fill=none,draw=none,at={(0.405,1.35)},anchor=north west},
            title style={font=\footnotesize},
            grid style={dotted,gray!50},
            width=\textwidth,
            height=0.25\textwidth,
            ybar,
            bar width=4pt,
            nodes near coords,
            every node near coord/.append style={font=\tiny,/pgf/number format/fixed, /pgf/number format/zerofill, /pgf/number format/precision=2, rotate=90, anchor=west}
        ]
            \addlegendentry{HTM}
            \addplot[color=darkcoral,pattern=north west lines,pattern color=darkcoral] coordinates {
                (1,  47.00) 
                (2,  47.40) 
                (3,  32.40) 
                (4,  0    ) 
                (5,  0    ) 
                (6,  49.90) 
                (7,  0    ) 
                (8,  99.30) 
                (9,  98.76) 
                (10, 17.51) 
            };
            \addlegendentry{Pot}
            \addplot[color=darkcyan,fill=darkcyan!30] coordinates {
                (1,  04.60) 
                (2,  0    ) 
                (3,  00.06) 
                (4,  0    ) 
                (5,  0    ) 
                (6,  49.85) 
                (7,  0    ) 
                (8,  00.01) 
                (9,  03.11) 
                (10, 17.39) 
            };
        \end{axis}
    \end{tikzpicture}
    \caption{
    Percentage of transactions that experience persistent aborts using baseline HTM transactions and Pot fast HTM transactions on the STAMP benchmarks (lower is better).
    $-$ and $+$ refer to the relative levels of contention in the configuration.
     }
    \label{fig:evaluation_htm_hwsuccess}
\end{figure}
While our Pot \acrshort{stm} fast transaction is able to reduce concurrency control overheads, implementing a \acrshort{htm} fast transaction that effectively reduces concurrency control overheads would require hardware support that is currently unavailable in existing processors.
However, by exploiting IBM's \acrlong{rot}s~(\acrshort{rot}s), Pot \acrshort{htm} fast transactions enjoy increased capacity limits, which increases the chance of committing more transactions entirely in hardware without falling back to the global lock.

We executed each benchmark with regular HTM and Pot using a single thread.
Since there is only one thread executing there are no aborts due to concurrency, however transactions may still abort for spuriously; thus we only count aborts that the hardware hints to be persistent---we collect this information from the TEXASR~register as we discuss in \S\ref{sub:implementation_htm}. 
Fig.~\ref{fig:evaluation_htm_hwsuccess} shows that the transactions that the baseline HTM cannot accommodate in both Labyrinth and Yada are also not accommodated by Pot's fast transaction.
The transactions of KMeans and SSCA2, on the other hand, can all execute without problem.
The rest of the benchmarks have a mix of transactions that can and cannot execute in hardware.
In these we can clearly see the benefit of Pot's fast transactions: for example, in Bayes around 47\% of the transactions can not be accomodated by the baseline HTM but this number falls to around 5\% with Pot.
Indeed, with Pot the number of transactions that are not accomodated by the hardware falls down from more than 30\% to less than 5\%. 
This means that Pot fast HTM transactions are successful at avoiding to fall back to the global lock.
Thus, fast transactions manage to regain some of the parallelism lost to ordered commits when the baseline \acrshort{htm} falls back to software.
\subsubsection{Performance} 
\label{subsub:evaluation_htm_perf}
\begin{figure*}[tb]
    \centering
    \begin{tikzpicture}
        \begin{axis}[
            xmin=0.5, xmax=10.5,
            xtick={1,2,3,4,5,6,7,8,9,10},
            xticklabels={Bayes,Genome,Intruder,KMeans$-$,KMeans$+$,Labyrinth,SSCA2,Vacation$-$,Vacation$+$,Yada},
            x tick label style={font=\footnotesize,rotate=90,anchor=east},
            x label style={font=\footnotesize},
            ymin=0, ymax=10,
            ytick={0,2,4,6,8,10},
            ymajorgrids=true,
            y tick label style={font=\footnotesize},
            ylabel near ticks,
            ylabel={Exec. time normalized \\ to HTM execution},
            ylabel style={align=center,font=\footnotesize},
            legend columns=0,
            legend style={font=\footnotesize,fill=none,draw=none,at={(0.81,1.35)},anchor=north east},
            title style={font=\footnotesize},
            grid style={dotted,gray!50},
            width=\textwidth,
            height=0.25\textwidth,
            ybar,
            bar width=4pt,
            nodes near coords,
            every node near coord/.append style={font=\tiny,/pgf/number format/fixed, /pgf/number format/zerofill, /pgf/number format/precision=2, rotate=90, anchor=west}
        ]
            \addlegendentry{2 threads}
            \addplot[color=darkcyan,pattern=north west lines,pattern color=darkcyan] coordinates {
                (1,  .81 ) 
                (2,  .93 ) 
                (3,  1.94) 
                (4,  1.10) 
                (5,  1.39) 
                (6,  .99 ) 
                (7,  1.31) 
                (8,  1.00) 
                (9,  1.03) 
                (10, 1.27) 
            };
            \addlegendentry{4 threads}
            \addplot[color=darkcyan,fill=darkcyan!30] coordinates {
                (1,  1.05) 
                (2,  .98 ) 
                (3,  1.92) 
                (4,  1.20) 
                (5,  1.70) 
                (6,  .99 ) 
                (7,  1.36) 
                (8,  1.05) 
                (9,  1.08) 
                (10, 1.26) 
            };
            \addlegendentry{8 threads}
            \addplot[color=darkcyan,pattern=north east lines,pattern color=darkcyan] coordinates {
                (1,  1.20) 
                (2,  .93 ) 
                (3,  1.96) 
                (4,  2.61) 
                (5,  6.46) 
                (6,  .99 ) 
                (7,  2.45) 
                (8,  1.03) 
                (9,  1.07) 
                (10, 1.33) 
            };
            \addlegendentry{16 threads}
            \addplot[color=darkcyan,pattern=crosshatch,pattern color=darkcyan] coordinates {
                (1,  1.35 ) 
                (2,  .94  ) 
                (3,  1.76 ) 
                (4,  5.43 ) 
                (5,  5.18 ) 
                (6,  .89  ) 
                (7,  2.64 ) 
                (8,  .96  ) 
                (9,  1.02 ) 
                (10, 1.29 ) 
            };
        \end{axis}
    \end{tikzpicture}
    \caption{
    Deterministic execution of STAMP using Pot.
    %
    %
    The y axis measures the execution time normalized to the nondeterministic execution using the baseline \acrshort{htm} (lower is better).
    $-$ and $+$ refer to the relative levels of contention in the configuration.
     }
    \label{fig:evaluation_htm_perf}
\end{figure*}
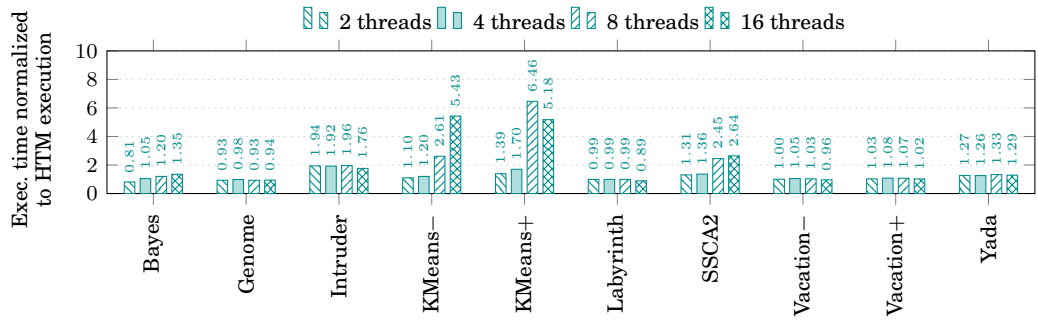
Fig.~\ref{fig:evaluation_htm_perf} shows the overhead of deterministic multithreading using Pot \acrshort{htm}\@.
It has less overhead on benchmarks where the baseline often falls back to software~(Bayes, Genome, Vacation).
In Genome Pot always outperforms the nondeterministic execution.
Vacation's results in Fig.~\ref{fig:evaluation_htm_perf} may seem unintuitive given than the baseline HTM practically always falls back to the global lock while Pot mostly executes without resorting the global lock~(Fig.~\ref{fig:evaluation_htm_hwsuccess}).
However, note that since all transactions executing in speculative mode exceed the hardware capacity, Pot is also executing one transaction at a time, albeit in fast mode instead of needing to fall back to the global lock.  

Arguably the more interesting benchmarks are the ones where the baseline performs well, i.e. falls~back less to the global lock~(Intruder, KMeans, SSCA2, and Yada from Fig.~\ref{fig:evaluation_htm_hwsuccess}).
In Intruder and Yada, in Fig.~\ref{fig:evaluation_htm_perf}, Pot achieves modest overheads of up to~$2 \times$.
KMeans and SSCA2 are optimal for the baseline \acrshort{htm}, featuring small transactions with few accesses and conflicts.
These characteristics make it difficult to mask the overheads of ensuring determinism.
KMeans also features an abundant use of thread synchronization via barriers which amplifies the overhead caused by the sequencer while assigning sequence numbers deterministically.
Also note that since fast transactions are not sped up in \acrshort{htm}, there is a noticeable drop in performance from 4 to 8 threads and even more from 8 to 16 threads due to the increased memory latencies of the NUMA architecture and hardware oversubscribing.

To the best of our knowledge, Pot advances the state of the art by enabling deterministic execution of \acrshort{htm}-based multithreaded programs for the first time.
Overall, Pot achieves deterministic execution with lower overhead at lower thread counts, but increased memory latencies lead to a drop in performance relatively to the nondeterministic baseline.
Efficiently achieving deterministic execution in the presence of non-uniform memory accesses represents an interesting future research avenue.
The results achieved by Pot \acrshort{stm} fast transactions suggest that hardware support for fast transactions that do not abort due to conflicts with other transactions may be worthwhile. 
\section{Related work} 
\label{sec:related}
\noindent{\textbf{Preordered transactions.}} 
The idea of preordering transactions has also been used in the database community with the objective of reducing the costs of distributed transactions~\cite{calvin-sigmod-2012}.
In our work we advocate its use to achieve deterministic multithreading of \acrshort{tm}-based programs.
The aforementioned works also describe a concurrency control protocol to ensure that the predefined order is respect.
However, that protocol could not be used in the context of our work, particularly in \acrshort{htm}, because it is tailored for transactions whose read/write sets can be determined a~priori and is based on~\acrlong{2pl}.
Ours is based on \acrshort{occ} and more general because it fully supports both transactions with static and dynamic read/write~sets.

\noindent{\textbf{Deterministic multithreading.}} 
Many deterministic multithreading systems for lock-based programs have been proposed~\cite{coredet-asplos-2010,dthreads-sosp-2011,kendo-asplos-2009,rfdet-ppopp-2014,parrot-sosp-2013}.
If transactions are implemented with locks, deterministic transactions could be implemented using such systems.
However this approach cannot be applied to off-the-shelf \acrshort{htm}, because the concurrency control is implemented in the hardware, and fails to exploit the semantics of transactions to reduce the overhead of ensuring determinism, because determinism is enforced with locks, which are at a level of abstraction lower than transactions.
Both DeTrans~\cite{detrans-sbacpad-2014} and DeSTM~\cite{destm-pact-2014} adapt the double~barrier technique used by many deterministic lock-based systems to \acrshort{stm}.
DeTrans provides strong determinism~\cite{kendo-asplos-2009} while DeSTM, like Pot, provides weak determinism~\cite{kendo-asplos-2009}.
One of the key differences between DeSTM and Pot is that in Pot the sequencer establishes a deterministic transaction serialization order that is enforced, i.e., the final outcome is as if transactions executed in the serial order defined by the sequencer.
DeSTM, on the other hand, uses a token-passing scheme that defines a deterministic order in which threads attempt to commit transactions.
Thus, the final outcome is always equivalent to the same transaction serialization order, although that order is unknown beforehand.
As a consequence of this design, DeSTM orders both aborts and commits and requires conflicts to be deterministic.
Pot only orders commits, and works whether conflicts are deterministic or not.
Pot's design allows it to achieve better performance than DeSTM (which is important when replicating applications for fault tolerance), and from the application developers point of view, they are equally helpful with concurrency bugs such as atomicity and order violations.
Furthermore, Pot's sequencer enables more uses cases, e.g. record-replay.
However, if concurrency bugs lie in the STM implementation then DeSTM is better due to its requirement of deterministic conflicts.
%
Grace~\cite{grace-oopsla-2009} ensures deterministic execution of fork-join programs by using a custom~\acrshort{stm} that uses a technique similar to ordered~commits.
DMP-TM~\cite{dmp-asplos-2009} also ensures determinism using hardware transactions and ordered commits.
Pot works with both \acrshort{stm} and \acrshort{htm}, is not limited to fork-join-style parallelism, and takes advantage of the deterministic order to improve efficiency via fast transactions.
DMP-TMFwd~\cite{dmp-asplos-2009} proposes to speculatively forward values written from transactions to their successors in the order.
However current hardware does not support this functionality, and it is unclear how this can be done without violating opacity.

\noindent{\textbf{Parallelizing sequential code.}} 
FastPath~\cite{fastpath-lcpc-2009}, IPOT~\cite{ipot-ppopp-2007} and TEPO~\cite{tepo-taco-2014} present a programming model to parallelize sequential code, e.g. loops, using transactions.
Like DMP-TMFwd, they also order commits and propose to speculatively forward values.
IPOT relies on unavailable hardware support to do so, TEPO does not implement it, and neither discusses how to do it while perserving opacity.
Pot ensures deterministic execution of multithreaded~\acrshort{tm} programs, exploits the deterministic order using fast transactions, and preserves opacity.

\noindent{\textbf{Transaction modes.}} 
Executing transactions with the guarantee that they do not abort has been used to support I/O inside transactions~\cite{inevitability-icpp-2008,irrevocability-spaa-2008}, and to improve the performance of \acrshort{stm}s at low thread counts~\cite{fastlane-ppopp-2014}.
Pot fast transactions are similar in spirit to these works, however, their intent is to minimize the overhead of ensuring determinism.
Unlike the aforementioned works, multiple fast transactions can safely execute in parallel by exploiting the existence of a predefined serialization order, e.g. a string of successive transactions that do not have read-write nor write-write conflicts between them can execute simultaneously as fast transactions.
PhTM~\cite{phtm-transact-2007} can execute transactions in different modes but only one mode is active at a time, while Pot executes different transactions in different modes simultaneously and transactions can switch modes at runtime.
\section{Conclusions} 
\label{sec:conclusion}
We presented Pot, a system that uses the concept of preordered transactions as a principled approach to achieve deterministic multithreaded execution of transactions.
At Pot's core is a novel concurrency control protocol that efficiently enforces a predefined transaction serialization order using two techniques: ordered commit and transaction modes.
Pot advances the state of the art by: (1) to the best of our knowledge, enabling deterministic execution of off-the-shelf HTM-based multithreaded programs, and (2) clearly outperforming the state of the art in STM-based deterministic execution while simultaneously achieving determinism with low overhead, providing promising evidence that using both \acrshort{stm} and determinism to enable multithreaded replicas for fault tolerance, and/or ease multithreaded programming, may be practical.

\begin{acks}
	We are grateful to the anonymous reviewers for their helpful comments and suggestions, and to Koen De Bosschere for acting as a proxy for additional discussion with the reviewers. 
	We also thank Tomas Vojnar, Daniel Hor\'{a}k, Jakub Cajka, Jaromir Capik, and the folks at Red Hat in Brno, Czech Republic, for providing us access to the infrastructure used in the evaluation.
	This research is supported by SFRH/BD/84497/2012 and PEst/UID/CEC/04516/2013.
\end{acks}

\bibliographystyle{ACM-Reference-Format-Journals}
\bibliography{references}


\begin{thebibliography}{00}


\ifx \showCODEN    \undefined \def \showCODEN     #1{\unskip}     \fi
\ifx \showDOI      \undefined \def \showDOI       #1{{\tt DOI:}\penalty0{#1}\ }
  \fi
\ifx \showISBNx    \undefined \def \showISBNx     #1{\unskip}     \fi
\ifx \showISBNxiii \undefined \def \showISBNxiii  #1{\unskip}     \fi
\ifx \showISSN     \undefined \def \showISSN      #1{\unskip}     \fi
\ifx \showLCCN     \undefined \def \showLCCN      #1{\unskip}     \fi
\ifx \shownote     \undefined \def \shownote      #1{#1}          \fi
\ifx \showarticletitle \undefined \def \showarticletitle #1{#1}   \fi
\ifx \showURL      \undefined \def \showURL       #1{#1}          \fi

\bibitem[\protect\citeauthoryear{Bergan, Anderson, Devietti, Ceze, and
  Grossman}{Bergan et~al\mbox{.}}{2010}]%
        {coredet-asplos-2010}
{Tom Bergan}, {Owen Anderson}, {Joseph Devietti}, {Luis Ceze}, {and} {Dan
  Grossman}. 2010.
\newblock \showarticletitle{{CoreDet}: A compiler and runtime system for
  deterministic multithreaded execution}. In {\em International Conference on
  Architectural Support for Programming Languages and Operating Systems
  (ASPLOS)}.
\newblock
\showDOI{%
\url{http://dx.doi.org/10.1145/1736020.1736029}}


\bibitem[\protect\citeauthoryear{Berger, Yang, Liu, and Novark}{Berger
  et~al\mbox{.}}{2009}]%
        {grace-oopsla-2009}
{Emery~D. Berger}, {Ting Yang}, {Tongping Liu}, {and} {Gene Novark}. 2009.
\newblock \showarticletitle{Grace: Safe Multithreaded Programming for {C/C++}}.
  In {\em ACM SIGPLAN International Conference on Object-Oriented Programming,
  Systems, Languages, and Applications (OOPSLA)}.
\newblock
\showDOI{%
\url{http://dx.doi.org/10.1145/1640089.1640096}}


\bibitem[\protect\citeauthoryear{Bernstein, Hadzilacos, and Goodman}{Bernstein
  et~al\mbox{.}}{1987}]%
        {concurrencycontrol-1987}
{Philip Bernstein}, {Vassos Hadzilacos}, {and} {Nathan Goodman}. 1987.
\newblock {\em Concurrency control and recovery in database systems}.
\newblock Addison-Wesley.
\newblock


\bibitem[\protect\citeauthoryear{{C++ Committee SG5}}{{C++ Committee
  SG5}}{2015}]%
        {cpp-tm-iso}
{{C++ Committee SG5}}. 2015.
\newblock Technical Specification for {C++} Extensions for Transactional
  Memory.
\newblock open-std.org/jtc1/sc22/wg21/docs/papers/2015/n4514.pdf.   (2015).
\newblock


\bibitem[\protect\citeauthoryear{Cain, Michael, Frey, May, Williams, and
  Le}{Cain et~al\mbox{.}}{2013}]%
        {tm-powerpc-isca-2013}
{Harold Cain}, {Maged Michael}, {Brad Frey}, {Cathy May}, {Derek Williams},
  {and} {Hung Le}. 2013.
\newblock \showarticletitle{Robust architectural support for transactional
  memory in the {POWER} architecture}. In {\em International Symposium on
  Computer Architecture (ISCA)}.
\newblock
\showDOI{%
\url{http://dx.doi.org/10.1145/2485922.2485942}}


\bibitem[\protect\citeauthoryear{Cui, Simsa, Lin, Li, Blum, Xu, Yang, Gibson,
  and Bryant}{Cui et~al\mbox{.}}{2013}]%
        {parrot-sosp-2013}
{Heming Cui}, {Jiri Simsa}, {Y. Lin}, {Hao Li}, {Ben Blum}, {Xinan Xu},
  {Junfeng Yang}, {Garth Gibson}, {and} {Randal Bryant}. 2013.
\newblock \showarticletitle{Parrot: A practical runtime for deterministic,
  stable, and reliable threads}. In {\em ACM Symposium on Operating Systems
  Principles (SOSP)}.
\newblock
\showDOI{%
\url{http://dx.doi.org/10.1145/2517349.2522735}}


\bibitem[\protect\citeauthoryear{Dalessandro, Spear, and Scott}{Dalessandro
  et~al\mbox{.}}{2010}]%
        {norec-ppopp-2010}
{Luke Dalessandro}, {Michael~F. Spear}, {and} {Michael~L. Scott}. 2010.
\newblock \showarticletitle{{NOrec}: Streamlining {STM} by Abolishing Ownership
  Records}. In {\em ACM SIGPLAN Symposium on Principles and Practice of
  Parallel Programming (PPoPP)}.
\newblock
\showDOI{%
\url{http://dx.doi.org/10.1145/1693453.1693464}}


\bibitem[\protect\citeauthoryear{Devietti, Lucia, Ceze, and Oskin}{Devietti
  et~al\mbox{.}}{2009}]%
        {dmp-asplos-2009}
{Joseph Devietti}, {Brandon Lucia}, {Luis Ceze}, {and} {Mark Oskin}. 2009.
\newblock \showarticletitle{{DMP}: Deterministic Shared Memory
  Multiprocessing}. In {\em International Conference on Architectural Support
  for Programming Languages and Operating Systems (ASPLOS)}.
\newblock
\showDOI{%
\url{http://dx.doi.org/10.1145/1508244.1508255}}


\bibitem[\protect\citeauthoryear{Dice, Shalev, and Shavit}{Dice
  et~al\mbox{.}}{2006}]%
        {tl2-disc-2006}
{Dave Dice}, {Ori Shalev}, {and} {Nir Shavit}. 2006.
\newblock \showarticletitle{{Transactional locking II}}. In {\em International
  Symposium on Distributed Computing (DISC)}.
\newblock
\showDOI{%
\url{http://dx.doi.org/10.1007/11864219_14}}


\bibitem[\protect\citeauthoryear{{Free Software Foundation}}{{Free Software
  Foundation}}{2014}]%
        {gcc-tm}
{{Free Software Foundation}}. 2014.
\newblock Transactional memory in {GCC}.
\newblock {gcc.gnu.org/wiki/TransactionalMemory}.   (2014).
\newblock


\bibitem[\protect\citeauthoryear{Gonzalez-Mesa, Gutierrez, Zapata, and
  Plata}{Gonzalez-Mesa et~al\mbox{.}}{2014}]%
        {tepo-taco-2014}
{M.~A. Gonzalez-Mesa}, {Eladio Gutierrez}, {Emilio~L. Zapata}, {and} {Oscar
  Plata}. 2014.
\newblock \showarticletitle{Effective Transactional Memory Execution Management
  for Improved Concurrency}.
\newblock {\em ACM Transactions on Architecture and Code Optimization (TACO)\/}
  {11}, 3 (2014).
\newblock
\showDOI{%
\url{http://dx.doi.org/10.1145/2633048}}


\bibitem[\protect\citeauthoryear{Guerraoui and Kapalka}{Guerraoui and
  Kapalka}{2008}]%
        {opacity-ppopp-2008}
{Rachid Guerraoui} {and} {Michal Kapalka}. 2008.
\newblock \showarticletitle{On the Correctness of Transactional Memory}. In
  {\em ACM SIGPLAN Symposium on Principles and Practice of Parallel Programming
  (PPoPP)}.
\newblock
\showDOI{%
\url{http://dx.doi.org/10.1145/1345206.1345233}}


\bibitem[\protect\citeauthoryear{Guerraoui, Kapalka, and Vitek}{Guerraoui
  et~al\mbox{.}}{2007}]%
        {stmbench7-eurosys-2007}
{Rachid Guerraoui}, {Michal Kapalka}, {and} {Jan Vitek}. 2007.
\newblock \showarticletitle{{STMBench7}: A Benchmark for Software Transactional
  Memory}. In {\em ACM European Conference on Computer Systems (EuroSys)}.
\newblock
\showDOI{%
\url{http://dx.doi.org/10.1145/1272996.1273029}}


\bibitem[\protect\citeauthoryear{Herlihy and Moss}{Herlihy and Moss}{1993}]%
        {htm-isca-1993}
{Maurice Herlihy} {and} {J.~Eliot~B. Moss}. 1993.
\newblock \showarticletitle{Transactional memory: Architectural support for
  lock-free data structures}. In {\em International Symposium on Computer
  Architecture (ISCA)}.
\newblock
\showDOI{%
\url{http://dx.doi.org/10.1145/165123.165164}}


\bibitem[\protect\citeauthoryear{Kung and Robinson}{Kung and Robinson}{1981}]%
        {occ-1981}
{H. Kung} {and} {John Robinson}. 1981.
\newblock \showarticletitle{On optimistic methods for concurrency control}.
\newblock {\em ACM Transactions on Database Systems (TODS)\/} {6}, 2 (1981).
\newblock
\showDOI{%
\url{http://dx.doi.org/10.1145/319566.319567}}


\bibitem[\protect\citeauthoryear{Lev, Moir, and Nussbaum}{Lev
  et~al\mbox{.}}{2007}]%
        {phtm-transact-2007}
{Yossi Lev}, {Mark Moir}, {and} {Dan Nussbaum}. 2007.
\newblock \showarticletitle{{PhTM}: Phased transactional memory}. In {\em ACM
  SIGPLAN Workshop on Transactional Computing (TRANSACT)}.
\newblock


\bibitem[\protect\citeauthoryear{Liu, Curtsinger, and Berger}{Liu
  et~al\mbox{.}}{2011}]%
        {dthreads-sosp-2011}
{Tongping Liu}, {Charlie Curtsinger}, {and} {Emery Berger}. 2011.
\newblock \showarticletitle{DThreads: Efficient deterministic multithreading}.
  In {\em ACM Symposium on Operating Systems Principles (SOSP)}.
\newblock
\showDOI{%
\url{http://dx.doi.org/10.1145/2043556.2043587}}


\bibitem[\protect\citeauthoryear{Lu, Zhou, Bergan, and Wang}{Lu
  et~al\mbox{.}}{2014}]%
        {rfdet-ppopp-2014}
{Kai Lu}, {Xu Zhou}, {Tom Bergan}, {and} {Xiaoping Wang}. 2014.
\newblock \showarticletitle{Efficient Deterministic Multithreading Without
  Global Barriers}. In {\em ACM SIGPLAN Symposium on Principles and Practice of
  Parallel Programming (PPoPP)}.
\newblock
\showDOI{%
\url{http://dx.doi.org/10.1145/2555243.2555252}}


\bibitem[\protect\citeauthoryear{Lu, Park, Seo, and Zhou}{Lu
  et~al\mbox{.}}{2008}]%
        {concurrencybugs-asplos-2008}
{Shan Lu}, {Soyeon Park}, {Eunsoo Seo}, {and} {Yuanyuan Zhou}. 2008.
\newblock \showarticletitle{Learning from Mistakes: A Comprehensive Study on
  Real World Concurrency Bug Characteristics}. In {\em International Conference
  on Architectural Support for Programming Languages and Operating Systems
  (ASPLOS)}.
\newblock
\showDOI{%
\url{http://dx.doi.org/10.1145/1346281.1346323}}


\bibitem[\protect\citeauthoryear{Minh, Chung, Kozyrakis, and Olukotun}{Minh
  et~al\mbox{.}}{2008}]%
        {stamp-iiswc-2008}
{Ch\'{\i}~Cao Minh}, {JaeWoong Chung}, {Christos Kozyrakis}, {and} {Kunle
  Olukotun}. 2008.
\newblock \showarticletitle{{STAMP}: Stanford transactional applications for
  multi-processing}. In {\em IEEE International Symposium on Workload
  Characterization (IISWC)}.
\newblock
\showDOI{%
\url{http://dx.doi.org/10.1109/IISWC.2008.4636089}}


\bibitem[\protect\citeauthoryear{Olszewski, Ansel, and Amarasinghe}{Olszewski
  et~al\mbox{.}}{2009}]%
        {kendo-asplos-2009}
{Marek Olszewski}, {Jason Ansel}, {and} {Saman Amarasinghe}. 2009.
\newblock \showarticletitle{Kendo: Efficient deterministic multithreading in
  software}. In {\em International Conference on Architectural Support for
  Programming Languages and Operating Systems (ASPLOS)}.
\newblock
\showDOI{%
\url{http://dx.doi.org/10.1145/1508244.1508256}}


\bibitem[\protect\citeauthoryear{Ravichandran, Gavrilovska, and
  Pande}{Ravichandran et~al\mbox{.}}{2014}]%
        {destm-pact-2014}
{Kaushik Ravichandran}, {Ada Gavrilovska}, {and} {Santosh Pande}. 2014.
\newblock \showarticletitle{{DeSTM}: Harnessing Determinism in {STM}s for
  Application Development}. In {\em International Conference on Parallel
  Architectures and Compilation Techniques (PACT)}.
\newblock
\showDOI{%
\url{http://dx.doi.org/10.1145/2628071.2628094}}


\bibitem[\protect\citeauthoryear{Schneider}{Schneider}{1990}]%
        {smr-surveys-1990}
{Fred~B. Schneider}. 1990.
\newblock \showarticletitle{Implementing Fault-tolerant Services Using the
  State Machine Approach: A Tutorial}.
\newblock {\em ACM Computing Surveys (CSUR)\/} {22}, 4 (1990).
\newblock
\showDOI{%
\url{http://dx.doi.org/10.1145/98163.98167}}


\bibitem[\protect\citeauthoryear{Shavit and Touitou}{Shavit and
  Touitou}{1997}]%
        {stm-1997}
{Nir Shavit} {and} {Dan Touitou}. 1997.
\newblock \showarticletitle{Software transactional memory}.
\newblock {\em Distributed Computing\/} {10}, 2 (1997).
\newblock
\showDOI{%
\url{http://dx.doi.org/10.1007/s004460050028}}


\bibitem[\protect\citeauthoryear{Smiljkovic, Stipic, Fetzer, \"{U}nsal,
  Cristal, and Valero}{Smiljkovic et~al\mbox{.}}{2014}]%
        {detrans-sbacpad-2014}
{Vesna Smiljkovic}, {Srdan Stipic}, {Christof Fetzer}, {Osman \"{U}nsal},
  {Adri\'{a}n Cristal}, {and} {Mateo Valero}. 2014.
\newblock \showarticletitle{{DeTrans}: Deterministic and Parallel Execution of
  Transactions}. In {\em International Symposium on Computer Architecture and
  High Performance Computing (SBAC-PAD)}.
\newblock
\showDOI{%
\url{http://dx.doi.org/10.1109/SBAC-PAD.2014.20}}


\bibitem[\protect\citeauthoryear{Spear, Silverman, Dalessandro, Michael, and
  Scott}{Spear et~al\mbox{.}}{2008}]%
        {inevitability-icpp-2008}
{M.F. Spear}, {M. Silverman}, {L. Dalessandro}, {M.M. Michael}, {and} {M.L.
  Scott}. 2008.
\newblock \showarticletitle{Implementing and exploiting inevitability in
  software transactional memory}. In {\em International Conference on Parallel
  Processing (ICPP)}.
\newblock
\showDOI{%
\url{http://dx.doi.org/10.1109/ICPP.2008.55}}


\bibitem[\protect\citeauthoryear{Spear, Kelsey, Bai, Dalessandro, Scott, Ding,
  and Wu}{Spear et~al\mbox{.}}{2009}]%
        {fastpath-lcpc-2009}
{Michael~F Spear}, {Kirk Kelsey}, {Tongxin Bai}, {Luke Dalessandro}, {Michael~L
  Scott}, {Chen Ding}, {and} {Peng Wu}. 2009.
\newblock \showarticletitle{Fastpath speculative parallelization}. In {\em
  Workshop on Languages and Compilers for Parallel Computing}.
\newblock
\showDOI{%
\url{http://dx.doi.org/10.1007/978-3-642-13374-9_23}}


\bibitem[\protect\citeauthoryear{Thomson, Diamond, Weng, Ren, Shao, and
  Abadi}{Thomson et~al\mbox{.}}{2012}]%
        {calvin-sigmod-2012}
{Alexander Thomson}, {Thaddeus Diamond}, {Shu-Chun Weng}, {Kun Ren}, {Philip
  Shao}, {and} {Daniel Abadi}. 2012.
\newblock \showarticletitle{Calvin: Fast distributed transactions for
  partitioned database systems}. In {\em ACM International Conference on
  Management of Data (SIGMOD)}.
\newblock
\showDOI{%
\url{http://dx.doi.org/10.1145/2213836.2213838}}


\bibitem[\protect\citeauthoryear{von Praun, Ceze, and Ca\c{s}caval}{von Praun
  et~al\mbox{.}}{2007}]%
        {ipot-ppopp-2007}
{Christoph von Praun}, {Luis Ceze}, {and} {Calin Ca\c{s}caval}. 2007.
\newblock \showarticletitle{Implicit Parallelism with Ordered Transactions}. In
  {\em ACM SIGPLAN Symposium on Principles and Practice of Parallel Programming
  (PPoPP)}.
\newblock
\showDOI{%
\url{http://dx.doi.org/10.1145/1229428.1229443}}


\bibitem[\protect\citeauthoryear{Wamhoff, Fetzer, Felber, Rivi\`{e}re, and
  Muller}{Wamhoff et~al\mbox{.}}{2013}]%
        {fastlane-ppopp-2014}
{J. Wamhoff}, {Christof Fetzer}, {Pascal Felber}, {Etienne Rivi\`{e}re}, {and}
  {Gilles Muller}. 2013.
\newblock \showarticletitle{FastLane: Improving performance of software
  transactional memory for low thread counts}. In {\em ACM SIGPLAN Symposium on
  Principles and Practice of Parallel Programming (PPoPP)}.
\newblock
\showDOI{%
\url{http://dx.doi.org/10.1145/2442516.2442528}}


\bibitem[\protect\citeauthoryear{Welc, Saha, and Adl-Tabatabai}{Welc
  et~al\mbox{.}}{2008}]%
        {irrevocability-spaa-2008}
{Adam Welc}, {Bratin Saha}, {and} {A. Adl-Tabatabai}. 2008.
\newblock \showarticletitle{Irrevocable transactions and their applications}.
  In {\em ACM Symposium on Parallelism in Algorithms and Architectures (SPAA)}.
\newblock
\showDOI{%
\url{http://dx.doi.org/10.1145/1378533.1378584}}


\bibitem[\protect\citeauthoryear{Yoo, Hughes, Lai, and Rajwar}{Yoo
  et~al\mbox{.}}{2013}]%
        {tsx-sc-2013}
{Richard Yoo}, {Christopher Hughes}, {Konrad Lai}, {and} {Ravi Rajwar}. 2013.
\newblock \showarticletitle{Performance evaluation of {Intel®} transactional
  synchronization extensions for high-performance computing}. In {\em
  International Conference for High Performance Computing Networking, Storage,
  and Analysis (SC)}.
\newblock
\showDOI{%
\url{http://dx.doi.org/10.1145/2503210.2503232}}


\end{thebibliography}

\received{May 2016}{November 2016}{November 2016}

\end{document}